\documentclass[aps,prd,11pt,notitlepage,longbibliography,nofootinbib,tightenlines,preprintnumbers,superscriptaddress]{revtex4-2}

\usepackage{booktabs}
\usepackage{amsmath,amssymb,amsfonts,mathrsfs}
\usepackage{graphicx}
\usepackage{color}
\usepackage{tikz}
\usetikzlibrary{calc}
 \usetikzlibrary{decorations.text}
 \usetikzlibrary{decorations.pathreplacing}
 \usetikzlibrary{shapes}

\usepackage{dsfont}
\usepackage[pdftex]{hyperref}
\hypersetup{colorlinks=true, linkcolor=darkred, citecolor=blue, linktoc=page}
\definecolor{darkred}{rgb}{0.8,0.1,0.1}

\makeatletter\def\l@subsubsection#1#2{}%
\makeatother

\makeatletter
\def\p@subsection{}
\makeatother

\renewcommand{\thesection}{\arabic{section}}

\numberwithin{equation}{section}
\renewcommand\theequation{\arabic{section}.\arabic{equation}} 

\usepackage{subfigure}

\def\cA{{\cal A}}
\def\cB{{\cal B}}

\def\cD{{\cal D}}

\def\cG{{\cal G}}

\def\cM{{\cal M}}
\def\cN{{\cal N}}

\def\cW{{\cal W}}

\def\cZ{{\cal Z}}

\newcommand{\Nb}{\mathscr{N}}

\def\CC{\ensuremath{\mathds C}}
\def\RR{\ensuremath{\mathds R}}
\def\ZZ{\ensuremath{\mathds Z}}

\DeclareMathOperator{\vol}{vol}
\DeclareMathOperator{\Vol}{Vol}
\DeclareMathOperator{\sech}{sech}

\DeclareMathOperator{\Li}{Li}
\DeclareMathOperator{\csch}{csch}

\def\Im{\mathop{\rm Im}}
\def\Re{\mathop{\rm Re}}
\def\Res{\mathop{\rm Res}}

\newcommand{\nocontentsline}[3]{}
\newcommand{\tocless}[2]{\bgroup\let\addcontentsline=\nocontentsline#1{#2}\egroup}

\begin{document}

\title{3d defects in 5d: RG flows and defect F-maximization}

\author{Leonardo~Santilli} 
\email{santilli@tsinghua.edu.cn}

\affiliation{Yau Mathematical Sciences Center\\ Tsinghua University, Beijing, 100084, China\\[2mm]}

\author{Christoph F.~Uhlemann} 
\email{uhlemann@maths.ox.ac.uk}

\affiliation{Mathematical Institute, University of Oxford, \\
	Andrew-Wiles Building,  Woodstock Road, Oxford, OX2 6GG, UK\\[2mm]}

\begin{abstract}
We use a combination of  AdS/CFT and supersymmetric localization to study codimension-2 defects in 5d SCFTs and their gauge theory deformations. The 5d SCFTs are engineered by $(p,q)$ 5-brane webs, with defects realized by D3-branes ending on the 5-brane webs.
We obtain the defect free energies and find that gauge theory descriptions of the combined 5d/3d systems can be connected to the UV defect SCFTs through a form of F-maximization which extremizes over different gauge theory defects. This leads to a match between the defect free energies obtained from supersymmetric localization in the gauge theories on the one hand and string theory results on the other. We extend this match to defect RG flows.
\end{abstract}

\maketitle
\tableofcontents

\parskip 1mm

\section{Introduction and summary}

Defects are important objects in quantum field theory (QFT). They are non-local probes, famously indicate confinement, and lead to generalized symmetry structures. Defects in conformal field theories (CFTs) can either preserve part of the conformal symmetry or break it, leading to defect renormalization group (RG) flows. 
Various combinations of ambient CFT dimensions\footnote{We follow the common practice to  refer to the higher-dimensional CFT degrees of freedom as `ambient CFT' and to the lower-dimensional ones as defect, to reserve the term `bulk' for holographic discussions.} and defect dimensions have been studied extensively, e.g.\ Wilson loops, surface defects in 4d and 6d SCFTs and domain walls. Here we focus on 3d defects in 5d SCFTs, i.e.\ defects of codimension 2.

Among the possible combinations of defect and ambient CFT dimensions, 3d defects in 5d theories are a relatively little studied subject, with previous works including \cite{Gaiotto:2014ina,Benvenuti:2016wet,Ashok:2017lko,Nedelin:2017nsb,Nieri:2018ghd,Nieri:2018pev,Aprile:2018oau,Cheng:2021vtq}. One complication is that 5d ambient SCFTs are typically non-Lagrangian, though they may admit descriptions as gauge theories after turning on mass deformations. Many aspects of the 5d SCFTs can be studied using such gauge theory deformations, but to what extent this extends to defects is not a trivial question. 5d SCFTs can generally have multiple gauge theory descriptions capturing different aspects of the SCFTs emerging in the UV. The same can be expected to apply for 3d defects in 5d SCFTs.
In this work we will explore 3d defects in 5d SCFTs using string theory, and in 5d gauge theories using supersymmetric localization, with the goal to connect the two descriptions.

We consider 5d ambient SCFTs engineered by $(p,q)$ 5-brane webs in Type IIB string theory \cite{Aharony:1997ju,Aharony:1997bh}. Defects of codimension 2 in these theories can be realized in the brane construction by adding D3-branes which are orthogonal to the 5-brane web and end on it \cite{Gaiotto:2014ina}. A natural planar limit of 5d SCFTs engineered by brane webs amounts to scaling the charges of the external 5-branes of the brane web. Supergravity duals for general 5d SCFTs of this type were constructed in \cite{DHoker:2016ujz,DHoker:2016ysh,DHoker:2017mds,DHoker:2017zwj}. 
If the 5d SCFTs have relevant deformations flowing to gauge theories in the IR, supersymmetric localization can be used to access the 5d SCFTs emerging as UV fixed points. In the planar limit the gauge theories are ``long quiver theories", in which the number of gauge nodes as well as the ranks are large.
The matrix models arising from supersymmetric localization in such gauge theories were solved in the planar limit in \cite{Uhlemann:2019ypp}. We will use these string theory and field theory tools to study 3d defects in 5d QFTs. 
Extremization principles will play a role in both descriptions.

In the supergravity duals, whose geometry generally is $\rm AdS_6\times S^2\times \Sigma$, the 3d defects are described by probe D3-branes \cite{Gutperle:2020rty}. For conformal defects they wrap an $\rm AdS_4$ subspace in $\rm AdS_6$ and are localized at a fixed point in the internal space, while for massive defects the location of the D3-branes in the internal space depends on the $\rm AdS_4$ radial coordinate. Planar defects and the free energies for conformal defects were studied in \cite{Gutperle:2020rty}. 
We will formulate the results for conformal defects as an F-maximization involving only the 5-brane charges.
For non-conformal defects the choice of geometry within a conformal class makes a difference in terms of which subgroup of the conformal group is naturally preserved by relevant deformations. 
In this work we study defect RG flows for defects extending along an $S^3$ in $S^5$. This allows for comparison to field theory analyses using supersymmetric localization in gauge theory deformations of the SCFTs. 

For 3d defects described by 3d matter coupled to 5d (quiver) gauge theories, we first derive the general form of the defect contribution to the sphere partition function on $S^5$ using supersymmetric localization. We then obtain defect free energies in a sample of 5d theories in the planar limit and show that they reproduce the free energies for conformal defects  by a form of defect $F$-maximization, in which we extremize over the gauge node in the 5d quiver which the 3d defect is associated with. Finally, we incorporate relevant deformations of the conformal defects to trigger defect RG flows, obtain the free energies, and show that they also match the string theory results. 

An interesting general question in QFT is how to count degrees of freedom. Quantities which are monotonic along RG flows provide viable counting functions and have been established in various dimensions \cite{Zamolodchikov:1986gt,Komargodski:2011vj,Jafferis:2010un,Jafferis:2011zi,Klebanov:2011gs,Liu:2012eea,Casini:2012ei,Cordova:2015fha}. In 5d, the existence of a counting function remains conjectural, with supporting evidence in \cite{Jafferis:2012iv,Chang:2017cdx,Fluder:2020pym,Akhond:2022awd,Akhond:2022oaf}.
Whether RG monotones exist for the degrees of freedom associated with defects is a natural question.
For 1d defects in 2d CFTs the $g$-theorem \cite{Affleck:1991tk,Friedan:2003yc,Casini:2016fgb} provides an RG monotone, extended to arbitrary ambient dimensions in \cite{Cuomo:2021rkm}. 
For 2d and 4d defects, monotonicity of anomaly coefficients was established in \cite{Jensen:2015swa,Wang:2020xkc,Wang:2021mdq}, and proposals for general dimensions were discussed in \cite{Kobayashi:2018lil}.
For 3d defects, the defect contribution to the sphere free energy is a natural candidate, motivated by the 3d $F$-theorem \cite{Jafferis:2010un,Jafferis:2011zi,Klebanov:2011gs,Liu:2012eea,Casini:2012ei}.
Indeed, a defect $F$-theorem was recently established in \cite{Casini:2023kyj}. 

In this work we analyze the defect free energies for 3d defects in 5d, including defect RG Flows.
Our results add to the exactly known quantities for 5d SCFTs, in particular relating to defects, and further tighten the interplay between string theory and field theory methods for these theories. They extend studies of long quiver gauge theories and are consistent with the 3d defect $F$-theorem.

\medskip
{\bf Outline:} In sec.~\ref{sec:defects} we discuss 3d defects in 5d QFTs in string theory as D3-branes ending on 5-brane webs, and in field theory as 3d matter in 5d gauge theories.
In sec.~\ref{sec:holography} we analyze conformal D3-brane defects and defect RG flows for $S^3$ defects in 5d SCFTs on $S^5$ using AdS/CFT, and obtain the defect free energies. In sec.~\ref{sec:loc-tot} we discuss 3d defects in 5d gauge theories using supersymmetric localization. The general form of the defect partition function is derived in sec.~\ref{sec:localization}. We derive free energies for 3d defects in concrete 5d gauge theories in sec.~\ref{sec:localization-F}, \ref{sec:loc-gauge-sample}, and recover the string theory results for conformal defects in 5d SCFTs using a form of $F$-maximization in sec.~\ref{sec:CFE}. In sec.~\ref{sec:M1FE} we extend the match to defect RG flows.
In sec.~\ref{sec:Mtheory} we connect our discussion to M-theory.

\section{3d defects and RG flows in 5d SCFTs}\label{sec:defects}

In this section we first introduce 3d defects described by D3-branes ending on 5-brane webs and 5-brane junctions, following and extending \cite{Gaiotto:2014ina}. We start with conformal defects in 5d SCFTs (a 5-brane junction at a point with a D3-brane ending on the intersection point), and then discuss defect mass deformations and defect RG flows. In sec.~\ref{sec:gaugeanddefect} we discuss 3d defects in 5d gauge theories, described by 5-brane junctions resolved into 5-brane webs.

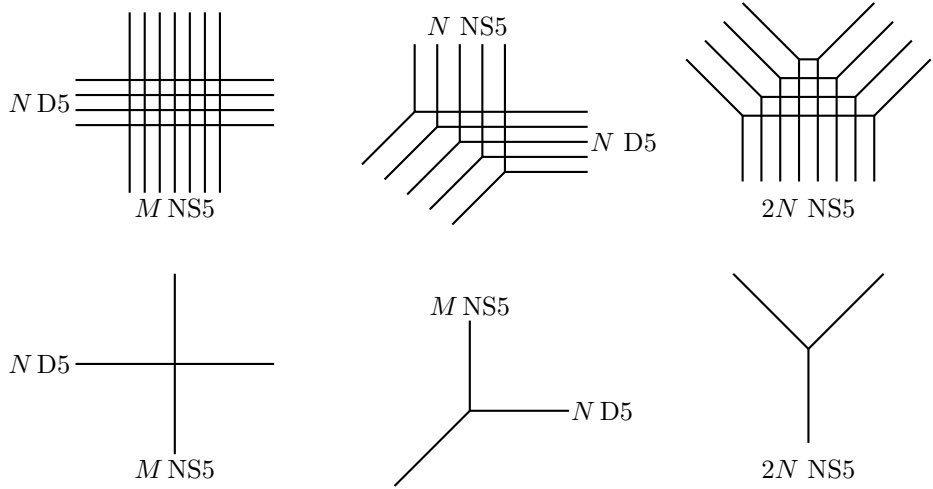
\begin{figure}
	\centering
	\begin{tabular}{ccc}
	\begin{tikzpicture}[scale=0.4]
		\foreach \i in {-1.5,-1.0,-0.5,0,0.5,1,1.5}{
			\draw[thick] (\i,-3) -- (\i,3);
		}
		\foreach \j in {-0.75,-0.25,0.25,0.75}{
			\draw[thick] (3.3,\j) -- (-3.3,\j);
		}		
		\node at (0,-3.5) {\small $M$\,NS5};
		\node at (-4.5,0) {\small $N$\,D5};
	\end{tikzpicture}
	&
	\hskip 10mm
	\begin{tikzpicture}[xscale=-0.4,yscale=-0.4]
		\draw[thick] (-4,0.75) -- (-0.5,0.75) -- (-0.5,-3);
		\draw[thick] (-0.5,0.75) -- +(1.75,1.75);
		
		\draw[thick] (-4,0.25) -- (0.25,0.25) -- (0.25,-3);
		\draw[thick] (0.25,0.25) -- +(1.75,1.75);
		
		\draw[thick] (-4,-0.25) -- (1.0,-0.25) -- (1.0,-3);
		\draw[thick] (1.0,-0.25) -- +(1.75,1.75);
		
		\draw[thick] (-4,-0.75) -- (1.75,-0.75) -- (1.75,-3);
		\draw[thick] (1.75,-0.75) -- +(1.75,1.75);
		
		\draw[thick] (-4,1.25) -- (-1.25,1.25) -- (-1.25,-3);
		\draw[thick] (-1.25,1.25) -- +(1.75,1.75);
		
		\node at (0,-3.6) {\small $N$ NS5};
		\node at (-5.2,0.25) {\small $N$ D5};
	\end{tikzpicture}
	\hskip 10mm
	&
	\begin{tikzpicture}[scale=0.5]
		\draw[thick] (-0.25,-2.5) -- (-0.25,0.75) -- (0.25,0.75) -- (0.25,-2.5);
		\draw[thick] (-0.25,0.75) -- +(-1.5,1.5);
		\draw[thick] (0.25,0.75) -- +(1.5,1.5);
		\draw[thick] (-0.75,-2.5) -- (-0.75,0.25) -- (0.75,0.25) -- (0.75,-2.5);
		\draw[thick] (-0.75,0.25) -- +(-1.5,1.5);
		\draw[thick] (0.75,0.25) -- +(1.5,1.5);
		\draw[thick] (-1.25,-2.5) -- (-1.25,-0.25) -- (1.25,-0.25) -- (1.25,-2.5);
		\draw[thick] (-1.25,-0.25) -- +(-1.5,1.5);
		\draw[thick] (1.25,-0.25) -- +(1.5,1.5);
		\draw[thick] (-1.75,-2.5) -- (-1.75,-0.75) -- (1.75,-0.75) -- (1.75,-2.5);
		\draw[thick] (-1.75,-0.75) -- +(-1.5,1.5);
		\draw[thick] (1.75,-0.75) -- +(1.5,1.5);
		
		\node at (0,-3.2) {$2N$ NS5};
	\end{tikzpicture}
	\\[5mm]
	\begin{tikzpicture}[scale=0.4]
		\draw[thick] (0,-3) -- (0,3);
		\draw[thick] (3.3,0) -- (-3.3,0);
		\node at (0,-3.5) {\small $M$\,NS5};
		\node at (-4.5,0) {\small $N$\,D5};
	\end{tikzpicture}
	&
	\hskip 10mm
	\begin{tikzpicture}[xscale=-0.4,yscale=-0.4]
		\draw[thick] (0,-3) -- (0,0) -- (-3.3,0);
		\draw[thick] (0,0) -- (2.5,2.5);
		
		\node at (0,-3.5) {\small $M$\,NS5};
		\node at (-4.4,0) {\small $N$\,D5};
	\end{tikzpicture}
	&
	\begin{tikzpicture}[scale=0.5]
		\draw[thick] (0,-2.5) -- (0,0) -- (2,2);
		\draw[thick] (0,0) -- (-2,2);
		
		\node at (0,-3.2) {$2N$ NS5};
	\end{tikzpicture}
	\end{tabular}
	\caption{Top: 5-brane webs engineering, from left to right, gauge theory deformations of the 5d $+_{N,M}$, $T_N$ and $Y_N$ SCFTs. Horizontal/vertical lines are D5/NS5 branes.
	The gauge theories are	in (\ref{eq:D5NS5-quiver}), (\ref{TNquiver}), (\ref{eq:YN-quiver}).
	Bottom: The corresponding UV fixed points are described by 5-brane junctions at a point.}
	\label{fig:pqwebs}
\end{figure}

We start with a $(p,q)$ 5-brane web giving rise to a 5d $\cN=1$ SCFT. We will focus on 3 concrete examples of 5d ambient SCFTs, which are the $T_N$ theories of \cite{Benini:2009gi,Hayashi:2014hfa} and the $+_{N,M}$ and $Y_N$ theories discussed in \cite{Bergman:2018hin}. The brane webs for the UV fixed points as well as for gauge theory deformations are shown in fig.~\ref{fig:pqwebs}.
Following \cite{Gaiotto:2014ina}, we introduce 3d defects described by D3-branes orthogonal to the plane of the 5-brane web. The orientations of the branes are: 
\begin{equation*}
	\begin{tabular}{l|c|c|c|c|c|c|c|c|c|c|}
		$ \ $ & $x^{0}$ & $x^{1}$& $x^{2}$& $x^{3}$& $x^{4}$& $x^{5}$& $x^{6}$& $x^{7}$& $x^{8}$& $x^{9}$ \\
		\hline
		D5 & $\bullet$ & $\bullet$ & $\bullet$  & $ \bullet $ & $\bullet$ & $\bullet$ & $ \ $ & $ \ $ & $ \ $ & $ \ $ \\
		\hline
		NS5 & $\bullet$ & $\bullet$ & $\bullet$ & $ \bullet $ & $\bullet$ & $ \ $ & $\bullet$ & $ \ $ & $ \ $ & $ \ $ \\
		\hline
		D3 & $\bullet$ & $\bullet$ & $\bullet$ & $ \ $ & $ \ $ & $ \ $  & $ \ $ &$ \bullet $ &$ \ $ & $ \ $ \\
		\hline
	\end{tabular}
\end{equation*}
Adding the D3-brane orthogonal to the $(x^5,x^6)$ plane breaks the $\mathfrak{so}(3) \cong \mathfrak{su}(2)_R $ rotation symmetry in $(x^7,x^8,x^9)$ to the $\mathfrak{so}(2)  \cong \mathfrak{u}(1)_{R}$ rotations transverse to the D3-brane, i.e. in the $(x^8,x^9)$-plane. Only half of the supercharges are preserved in presence of the D3-branes. 

\begin{figure}
	\centering
	\subfigure[][]{\label{fig:conf-D3-plus}
\begin{tikzpicture}[scale=0.7]
	
	\draw[blue] (-2,0) -- (-6,0);
	\draw[red] (-2,1) -- (-6,-1);
	\draw[black,thick] (-4,0) -- (-4,2);
	
	\begin{scope}[xshift=-2cm]
	\draw[->] (-6.5,1.5) -- (-6.5,2.3);
	\draw[->] (-6.5,1.5) -- (-5.7,1.5);
	\draw[->] (-6.5,1.5) -- (-5.9,1.8);
	\node[anchor=west] at (-5.7,1.5) {$\scriptstyle x^5$};
	\node[anchor=south west] at (-5.9,1.8) {$\scriptstyle x^6$};
	\node[anchor=south] at (-6.5,2.3) {$\scriptstyle x^7$};
	\end{scope}
	
	\node[anchor=east] at (-6,0) {\footnotesize $N$ D5};
	\node[anchor=east] at (-4,2) {\footnotesize D3};
	\node[anchor=north] at (-6,-1) {\footnotesize $M$ NS5};
\end{tikzpicture}
	}
\hskip 10mm
	\subfigure[][]{\label{fig:massive-D3-plus}
	\begin{tikzpicture}[scale=0.7]
		
		\draw[blue] (-2,0) -- (-6,0);
		\draw[red] (-2,1) -- (-2.94,0.53);
		\draw[red] (-3.06,0.47) -- (-6,-1);
		\draw[black,thick] (-3,0) -- (-3,2);
				
		\node[anchor=east] at (-6,0) {\footnotesize $N$ D5};
		\node[anchor=east] at (-3,2) {\footnotesize D3};
		\node[anchor=north] at (-6,-1) {\footnotesize $M$ NS5};
	\end{tikzpicture}
}
\hskip 10mm
\subfigure[][]{\label{fig:massive-D3-massive-5d}
	\begin{tikzpicture}[scale=0.7]
		\foreach \i in {-0.2,-0.1,0,0.1,0.2}{
		\draw[blue] (2+3*\i,\i) -- (6+3*\i,\i);
		}
		\draw[red] (1.7,-1) -- (5.7,1);
		\draw[red] (2,-1) -- (6,1);
		\draw[red] (2.3,-1) -- (6.3,1);
		\draw[black,thick] (4,0) -- (4,2);
		
		\node[anchor=east] at (1.7,0) {\footnotesize $N$ D5};
		\node[anchor=east] at (4,2) {\footnotesize D3};
		\node[anchor=north] at (2,-1) {\footnotesize $M$ NS5};
	\end{tikzpicture}
}

\caption{(a) conformal D3-brane defect in the $+_{N,M}$ SCFT. (b) defect mass deformation leading to a defect RG flow. (c) D3-brane defect in a gauge theory deformation of the 5d SCFT.\label{eq:D3-defect-5-brane}}
\end{figure}
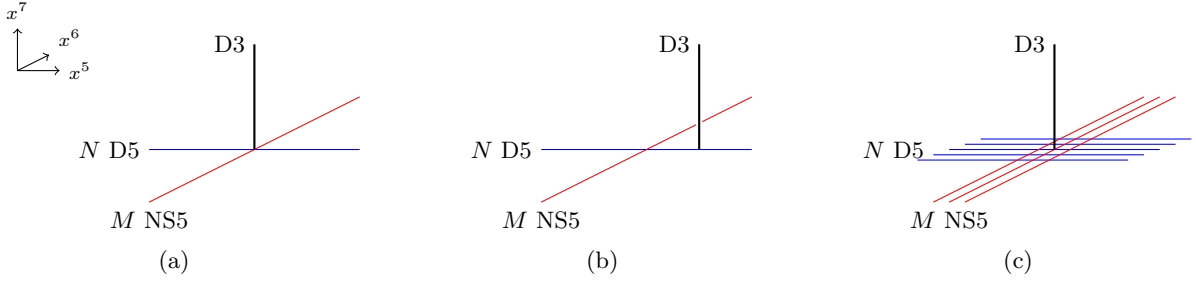

To realize conformal defects, we have to start with the ambient 5d SCFT at the fixed point, i.e.\ with a 5-brane junction at a point. To preserve conformal symmetry, the D3-brane then has to end on the intersection point. This is shown for the $+_{N,M}$ theory in fig.~\ref{fig:conf-D3-plus}. 
Since the D3 and D5/NS5 branes are not in Hanany--Witten orientations relative to each other \cite{Hanany:1996ie}, there is no distinction which D5 or NS5 brane the D3-brane ends on when the brane junction is resolved; there is only one conformal 3d defect.\footnote{This is different from the D3-branes discussed in \cite{Uhlemann:2020bek} which describe Wilson loops and are in a Hanany--Witten orientation w.r.t.\ the 5-branes. The UV fixed point then depends on where the D3 is in the resolved 5-brane web.}
The 5d SCFT without defect has conformal and R-symmetry algebra $\mathfrak{so}(2,5) \oplus \mathfrak{su}(2)_R$. The insertion of the defect breaks this symmetry,
\begin{align}
	\mathfrak{so}(2,5) \oplus \mathfrak{su}(2)_R \quad \longrightarrow\quad \mathfrak{so}(2,3) \oplus \mathfrak{u}(1)_{\perp } \oplus \mathfrak{u}(1)_{R}~.
\end{align}
These are, respectively the 3d conformal group, the rotations transverse to the defect, and the Cartan subalgebra $\mathfrak{u}(1)_{R} \subset \mathfrak{su}(2)_{R}$ of the 5d R-symmetry. From the point of view of the 3d theory, $ \mathfrak{u}(1)_{\perp } \oplus \mathfrak{u}(1)_{R}$ mix according to 
\begin{align}
	\mathfrak{u}(1)_{\perp } \oplus \mathfrak{u}(1)_{R} \quad \longrightarrow\quad \mathfrak{u}(1)_{f } \oplus \mathfrak{u}(1)_{r}~,
\end{align}
where $\mathfrak{u}(1)_r$ is the diagonal combination of $\mathfrak{u}(1)_{R}$ and $\mathfrak{u}(1)_{\perp} $ forming the 3d R-symmetry. The other independent combination is a flavor symmetry $\mathfrak{u}(1)_{f}$.

Defect mass deformations, i.e.\ relevant deformations on the defect, can be realized by displacing the D3-brane from the intersection point of the 5-brane junction. This is illustrated for the $+_{N,M}$ theory in fig.~\ref{fig:massive-D3-plus}. When the 5d SCFT is at the fixed point, there are only two defect mass deformations for a single D3: along the D5 and along the NS5. This type of defect mass deformation and the resulting defect RG flows will be studied quantitatively using AdS/CFT in sec.~\ref{sec:holography}.

The available defect mass deformations are different when the 5d theory is at a generic point of its (extended) moduli space. The D3 may be moved along the entire 5-brane web, and different parts of that moduli space may have different gauge theory descriptions. For example, fig.~\ref{fig:massive-D3-massive-5d} shows a deformation of the $+_{N,M}$ theory described by the gauge theory in (\ref{eq:D5NS5-quiver}). Depending on where the D3-brane is placed, it may describe 3d defects associated with different gauge nodes in the 5d quiver gauge theory. 
This leads to an interesting question, raised already in  \cite[sec.~6]{Gaiotto:2014ina}: 
\begin{itemize}
	\item[--] To what extent do D3-brane defects in interacting 5d SCFTs with gauge theory deformations admit a useful Lagrangian description as 3d matter coupled to the 5d gauge theory?
\end{itemize}
In a gauge theory deformation of a 5d SCFT,  conformal symmetry is already broken in the 5d theory, and there is no distinguished notion of a conformal defect. We can nevertheless consider the UV limit of the combined system, in which the brane configuration becomes scale-invariant, and hope to interpret this as the fixed point of the combined 5d/3d theory.
To investigate the extent to which this is possible we will complement brane engineering and AdS/CFT with localization calculations in the gauge theories and connect these descriptions.

In the following we first discuss defect theories and their moduli spaces for simple examples where the 5d theory is free and therefore has a Lagrangian description, before switching perspectives and discussing 3d defects in more general 5d gauge theories from the field theory standpoint.

\textbf{\boldmath{$+_{1,1}$} theory:} We start with a semi-infinite D3-brane placed at the intersection of a D5 and an NS5-brane, i.e.\ the $+_{1,1}$ theory. This case was discussed in In \cite[sec.~3]{Gaiotto:2014ina}.
The ambient 5d field theory is a free hypermultiplet. 
For the field theory descriptions for the combined 3d/5d system discussed in \cite{Gaiotto:2014ina} the D3-brane is terminated on a separate 5-brane.

\begin{figure}
	\centering
	\subfigure[][]{\label{fig:freehyp-0}
		\begin{tikzpicture}[scale=0.7]
			\begin{scope}[xshift=-1cm]
				\draw[->] (-3,0.5) -- (-3,1.3);
				\draw[->] (-3,0.5) -- (-2.2,0.5);
				\draw[->] (-3,0.5) -- (-2.4,0.8);
				\node[anchor=west] at (-2.3,0.5) {$\scriptstyle x^5$};
				\node[anchor=south west] at (-2.5,0.7) {$\scriptstyle x^6$};
				\node[anchor=south] at (-3,1.3) {$\scriptstyle x^7$};
			\end{scope}
			
			\draw[blue] (-2,0) -- (2,0);
			\draw[red] (-2,-1) -- (2,1);
			\draw[black,thick] (0,0) -- (0,1.5);
			\draw[blue] (-2,1.5) -- (2,1.5);
			
		\end{tikzpicture}
	}\hskip 10mm
	\subfigure[][]{\label{fig:freehyp-1}
		\begin{tikzpicture}[scale=0.7]
			\draw[blue] (-2,0) -- (2,0);
			\draw[red] (-2,-1) -- (2,1);
			\draw[black,thick] (1,0.5) -- (1,1.7);
			\draw[blue] (-1.5,1.7) -- (1.8,1.7);
			
		\end{tikzpicture}
	}\hskip 10mm
	\subfigure[][]{\label{fig:freehyp-2}
		\begin{tikzpicture}[scale=0.7]
			\draw[blue] (-2,0) -- (2,0);
			\draw[red] (-2,-1) -- (2,1);
			\draw[black,thick] (-1,0) -- (-1,1.5);
			\draw[blue] (-2,1.5) -- (2,1.5);
			
		\end{tikzpicture}
	}
	\caption{(a) 5d free hypermultiplet coupled to a conformal 3d defect, comprising a massless pair of 3d chiral/antichiral multiplets. (b) and (c): defect relevant deformations.}
	\label{fig:freehyp-various}
\end{figure}
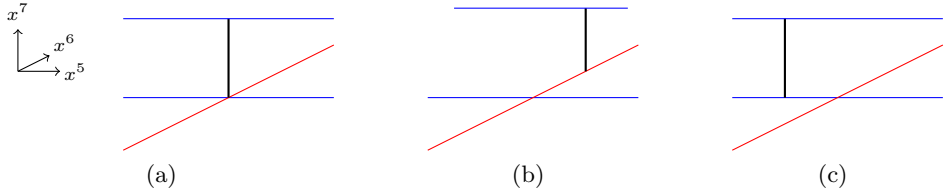

One option is to terminate the D3-brane on a D5-brane separated from the 5-brane web in the $x^7$ direction (fig.~\ref{fig:freehyp-0}). We denote this additional D5-brane as D5$_0$. The D3-D5 strings connecting the D3 to the $+_{1,1}$ web give rise to a pair of 3d chiral/antichiral multiplets with a $\mathfrak{u}(1)$ flavor symmetry.
When the ambient 5d hypermultiplet is at the conformal point (massless), two deformations are available. One corresponds to sliding the D3-brane along the NS5-brane, dragging the D5$_0$ along (fig.~\ref{fig:freehyp-1}). This deformation gives a real mass to the 3d chiral/antichiral pair. The other corresponds to moving the D3-brane along the D5 brane (fig.~\ref{fig:freehyp-2}). From the defect point of view, this is a Higgs branch deformation, and does not involve 5-brane motions.

We can also study the S-dual setup, with D5 and NS5-branes exchanged. This leads to a D3-brane suspended between the $+_{1,1}$ junction and an additional separated NS5$_0$ brane. We recognize a 3d $U(1)$ gauge theory with a chiral/antichiral pair. The S-dual of fig.~\ref{fig:freehyp-2} describes this gauge theory on the Coulomb branch. The S-dual of fig.~\ref{fig:freehyp-1} has the vertically separated NS5$_0$-brane dragged away. This corresponds to turning on a twisted mass for the topological $\mathfrak{u}(1)$ symmetry on the defect, which is a 3d FI parameter.

The setups with D3-branes terminating on additional 5-branes separated in the $x^7$ direction naturally arise from Higgs-branch flows of an enlarged 5d theory.
We could also terminate the D3-brane on a second $+_{1,1}$ web, separated from the first in the $x^7$ direction. In that case displacements of the D3-brane along the D5 or NS5 branes of the $+_{1,1}$ junction are both possible without moving 5-branes and the two are on equal footing. Comparing to the setup with a semi-infinite D3-brane as in fig.~\ref{eq:D3-defect-5-brane}, where movements along the D5 and NS5 branes are also on equal footing, we expect the moves in the latter to correspond to mass deformations from the 3d perspective.

\textbf{\boldmath{$+_{N,1}$} theory:} As a slight generalization we discuss the $+_{N,1}$ theory, describing $N^2$ free 5d hypermultiplets. Adding a D3-brane stretching between the intersection point of the $+_{N,1}$ junction and a separated D5$_0$, as in fig.~\ref{fig:freehyp-0}, now adds $N$ 3d chiral and antichiral multiplets. When the 5-branes all intersect at a point, corresponding to massless hypermultiplets, the defect mass deformations are the same as for the $+_{1,1}$ theory before. The deformation in fig.~\ref{fig:freehyp-1}, for example, corresponds to a homogeneous real mass for the 3d modes.

\begin{figure}
	\centering
	\subfigure[][]{\label{fig:PN1-2}
		\begin{tikzpicture}[scale=0.7]
			\draw[blue] (-2,0) -- (2,0);
			\draw[blue] (-1.9,0.05) -- (2.1,0.05);
			\draw[blue] (-1.8,0.1) -- (2.2,0.1);
			\draw[blue] (-1.7,0.15) -- (2.3,0.15);
			\draw[red] (-2,-1) -- (2,1);
			\draw[black,thick] (-1,0.1) -- (-1,1.3);
			\draw[blue] (-1.8,1.3) -- (1.5,1.3);
			
			\draw[->] (-3,0.5) -- (-3,1.3);
			\draw[->] (-3,0.5) -- (-2.2,0.5);
			\draw[->] (-3,0.5) -- (-2.4,0.8);
			\node[anchor=west] at (-2.3,0.5) {$\scriptstyle x^5$};
			\node[anchor=south west] at (-2.5,0.7) {$\scriptstyle x^6$};
			\node[anchor=south] at (-3,1.3) {$\scriptstyle x^7$};
		\end{tikzpicture}
	}\hskip 7mm
	\subfigure[][]{\label{fig:PN1-3}
		\begin{tikzpicture}[scale=0.7]
			\draw[blue] (-2,0) -- (2,0);
			\draw[red] (-1.8,-1) -- (2.2,1);
			\draw[red] (-1.9,-1) -- (2.1,1);
			\draw[red] (-2,-1) -- (2,1);
			\draw[red] (-2.1,-1) -- (1.9,1);
			\draw[black,thick] (1,0.5) -- (1,1.7);
			\draw[red] (-1,0.7) -- (1.6,2);
			
		\end{tikzpicture}
	}\hskip 7mm
	\subfigure[][]{\label{fig:PN1-4}
		\begin{tikzpicture}[scale=0.7]
			\draw[blue] (-2,0) -- (2,0);
			
			\draw[red] (-1.8,-1) -- (2.2,1);
			\draw[red] (-1.9,-1) -- (2.1,1);
			\draw[red] (-2,-1) -- (2,1);
			\draw[red] (-2.1,-1) -- (1.9,1);
			\draw[black,thick] (-1,0) -- (-1,1.3);
			\draw[red] (-2,0.8) -- (0.4,2);
			
		\end{tikzpicture}
	}
	\caption{(a) Defect Higgs branch configuration for $+_{N,1}$. (b) Defect configuration obtained from S-duality on (a). (c) Different massive deformation of the defect in (b).}
	\label{fig:PN1-various}
\end{figure}
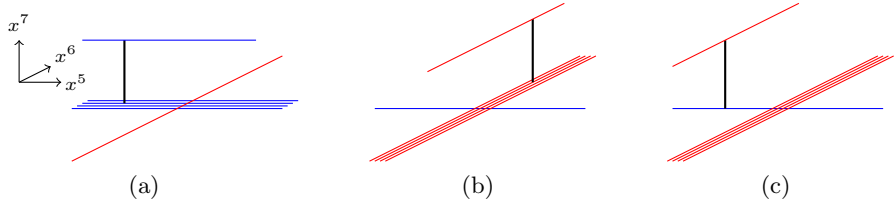\par

When the 5-brane junction is (partly) resolved, there are now $N$ inequivalent ways to suspend the D3-brane between the D5$_0$ and a D5-brane and move it in the $x^5$ direction (fig.~\ref{fig:PN1-2}). These are $N$ Higgs branch directions. When the 5-brane junction approaches the UV fixed point, these Higgs branches coalesce, otherwise they are spatially separated in the $x^6$ direction. 

The S-dual brane configuration is a D3-brane suspended between two NS5-branes (fig.~\ref{fig:PN1-3}), namely NS5$_j$ in the 5-brane web and NS5$_0$. This configuration describes a 3d $\mathcal{N}=2$ $U(1)_{(j)}$ gauge theory with one chiral/anitchiral pair on its Coulomb branch. For every Higgs branch deformation of the initial setup we obtain a distinct $U(1)_{(j)}$ gauge theory by S-duality. Sliding the D3-brane to the junction, we can again move it onto the D5-brane and in the $x^5$ direction (fig.~\ref{fig:PN1-4}). This turns on an FI parameter for the $U(1)_{(j)}$ gauge group.

\subsection{3d defects in 5d gauge theories}\label{sec:gaugeanddefect}

\begin{figure}
	\centering
	\begin{tikzpicture}[auto,square/.style={regular polygon,regular polygon sides=4}]
		\node[circle,draw] (gauge1) at (4.5,0) {$N_{t_0}$};
		\node[draw=none] (gaugemid) at (3,0) {$\cdots$};
		\node[draw=none] (gaugex) at (6,0) {$\cdots$};
		\node[circle,draw] (gauge0) at (7.5,0) {$N_{L}$};
		\node[circle,draw] (gauge3) at (1.5,0) {$N_2$};
		\node[circle,draw] (gauge4) at (0,0) {$N_1$};
		\node[square,draw] (fl1) at (4.5,-1.5) { \hspace{8pt} };
		\node[square,draw] (fl2) at (0,-1.5) { \hspace{8pt} };
		\node[square,draw] (fl3) at (1.5,-1.5) { \hspace{8pt} };
		\node[square,draw] (fl0) at (7.5,-1.5) { \hspace{8pt} };
		\node[draw=none] (aux1) at (4.5,-1.5) {$k_{t_0}$};
		\node[draw=none] (aux2) at (0,-1.5) {$k_1$};
		\node[draw=none] (aux3) at (1.5,-1.5) {$k_2$};
		\node[draw=none] (aux3) at (7.5,-1.5) {$k_L$};
		\node[square,draw=black,fill=gray,fill opacity=0.35] (flup) at (4.5,1.5) { \hspace{8pt} };
		\node[draw=none] (auxup) at (4.5,1.5) {$1$};
		\draw[-](gauge1)--(gaugemid);
		\draw[-](gaugemid)--(gauge3);
		\draw[-](gauge3)--(gauge4);
		\draw[-](gauge1)--(fl1);
		\draw[-](gauge4)--(fl2);
		\draw[-](gauge3)--(fl3);
		\draw[-](gauge0)--(fl0);
		\draw[-](gauge0)--(gaugex);
		\draw[-](gaugex)--(gauge1);
		\path[->] (gauge1) edge[bend left=30] node {$q$} (flup);
		\path[->] (flup) edge[bend left=30] node {$\tilde{q}$} (gauge1);
	\end{tikzpicture}
	\caption{Linear quiver with circular nodes denoting $SU(N_t)$ gauge groups and square nodes denoting $U(k_t)$ flavor groups. The defect $\cD$ (gray) consists of the 3d chiral/antichiral pair $(q,\tilde{q})$ charged under the 5d gauge node $SU(N_{t_0})$.}
	\label{fig:quiver}
\end{figure}
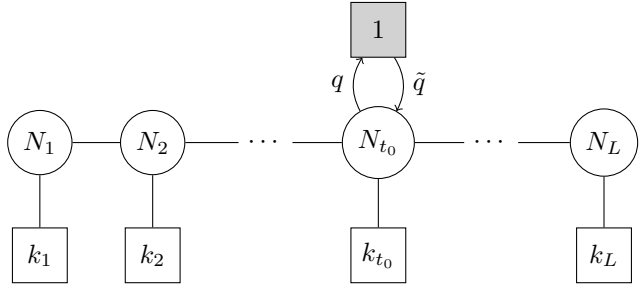

We will complement the string theory analyses in sec.~\ref{sec:holography} with field theory calculations using supersymmetric localization for  3d defects in 5d gauge theories. The theories engineered by the 5-brane webs shown in fig.~\ref{fig:pqwebs} have relevant deformations which are described by linear quiver gauge theories. 
The general form is shown in fig.~\ref{fig:quiver}: the gauge group consists of $SU(N_t)$ gauge nodes, while the matter content consists of bifundamental hypermultiplets between adjacent nodes and $k_t$ additional hypermultiplets in the fundamental of $SU(N_t)$, for $t=1,\ldots,L$.

Inspired by the discussion for free field theories above, we add to the 5d $\cN=1$ gauge theory a pair of 3d chiral/antichiral multiplets localized on a 3d submanifold. The resulting codimension-2 defect is denoted $\cD$. The inclusion of the defect preserves half of the ambient supersymmetry, yielding a 3d $\cN=2$ defect theory. 
The 3d fields are coupled to the ambient 5d theory by picking a gauge node $t_0 \in \left\{ 1, \dots, L \right\}$ and coupling the 3d chiral/antichiral pair $(q,\tilde{q})$ to the $SU(N_{t_0})$ vector multiplet, as shown in fig.~\ref{fig:quiver}.
The 3d (anti)chiral multiplets transform in the (anti)fundamental representation of the 5d $SU(N_{t_0})$ gauge group. The defect free energy will be discussed in sec.~\ref{sec:loc-tot}; it will be sensitive to the structure of the entire 5d quiver, even though the 3d fields are only coupled directly to one gauge node.

\par

At generic points on the Coulomb branch of the 5d quiver theory, the defect chiral/antichiral pair is subject to a generic vector mass matrix, and the defect moduli space is the disjoint union of one-quaternionic dimensional Higgs branches. Hence, when the 5d quiver theory is on its Coulomb branch, $\cD$ admits one real mass deformation, that turns on a background vector multiplet for the vector $\mathfrak{u}(1)$ flavor symmetry on the defect, and $N_{t_0}$ Higgs branch deformations.

The free energies for defects on an $S^3$ in an $S^5$ ambient space will be obtained using supersymmetric localization, and we will compare them to the string theory results.
To this end, the results are (implicitly) extrapolated to the UV fixed point of the 5d gauge theories. This is the opposite limit compared to the flows to the IR usually considered in 3d. One may therefore wonder about a potential role of  $F$-maximization in determining the proper R-symmetry. The 5d SCFT has an $\mathfrak{su}(2)_R$ R-symmetry, of which the combined 5d/3d system preserves a $\mathfrak{u}(1)_r$. For the class of defects we work out explicitly, we do not find it necessary to extremize over the R-symmetry as we move into the UV. However, when considering different 3d field theory descriptions for the defects, we find that the proper UV limit is obtained by extremizing $F$ over the gauge node which the defect is associated with. This will be described in detail in sec.~\ref{sec:M1FE}.

\section{D3-brane defects in \texorpdfstring{A\lowercase{d}S$_6$/CFT$_5$}{AdS6/CFT5}}\label{sec:holography}

The discussion of 3d defects in AdS$_6$/CFT$_5$ will be based on the Type IIB supergravity solutions describing $(p,q)$ 5-brane junctions, constructed in \cite{DHoker:2016ujz,DHoker:2016ysh,DHoker:2017mds,DHoker:2017zwj}.
The geometry is a warped product of $\rm AdS_6$ and $S^2$ over a Riemann surface $\Sigma$ with boundary, with the $S^2$ collapsing on the boundary of $\Sigma$.
The solutions are defined in terms of holomorphic functions $\cA_\pm$ on $\Sigma$. 
The Einstein-frame metric, complex two-form $C_{(2)}$, and axion-dilaton scalar $B=(1+i\tau)/(1-i\tau)$ are
\begin{align}\label{eqn:ansatz}
	ds^2 &= f_6^2 \, ds^2 _{\mathrm{AdS}_6} + f_2^2 \, ds^2 _{\mathrm{S}^2} 
	+ 4\rho^2\, |dw|^2~,
	\qquad\quad
	B =\frac{\partial_w \cA_+ \,  \partial_{\bar w} \cG - R \, \partial_{\bar w} \bar \cA_-   \partial_w \cG}{
		R \, \partial_{\bar w}  \bar \cA_+ \partial_w \cG - \partial_w \cA_- \partial_{\bar w}  \cG}~,
	\nonumber\\
	C_{(2)}&=\frac{2i}{3}\left(
	\frac{\partial_{\bar w}\cG\partial_w\cA_++\partial_w \cG \partial_{\bar w}\bar\cA_-}{3\kappa^{2}T^2} - \bar{\mathcal{A}}_{-} - \mathcal{A}_{+}  \right)\vol_{S^2}~,
\end{align}
where 
$w$ is a complex coordinate on $\Sigma$ and $ds^2_{\rm AdS_6}$ and $ds^2_{S^2}$ are the line elements for unit-radius $\rm AdS_6$ and $S^2$, respectively.
The metric functions are
\begin{align}\label{eq:metric-functions}
	f_6^2&=\sqrt{6\cG T}~, & f_2^2&=\frac{1}{9}\sqrt{6\cG}\,T ^{-\tfrac{3}{2}}~, & \rho^2&=\frac{\kappa^2}{\sqrt{6\cG}} T^{\tfrac{1}{2}}~,
\end{align}
where
\begin{align}\label{eq:kappa2-G}
	\kappa^2&=-|\partial_w \cA_+|^2+|\partial_w \cA_-|^2~,
	&
	\partial_w\cB&=\cA_+\partial_w \cA_- - \cA_-\partial_w\cA_+~,
	\nonumber\\
	\cG&=|\cA_+|^2-|\cA_-|^2+\cB+\bar{\cB}~,
	&
	T^2&=\left(\frac{1+R}{1-R}\right)^2=1+\frac{2|\partial_w\cG|^2}{3\kappa^2 \, \cG }~.
\end{align}
The differentials $\partial_w\cA_\pm$ have poles at points $r_\ell$ on the boundary of $\Sigma$, at which 5-branes with charge $(p_\ell,q_\ell)$ emerge. The charges $(p_\ell,q_\ell)$ are given by 
\begin{align}\label{eq:charges}
	\Res_{w=r_\ell} \partial_w\cA_\pm & = \frac{3}{4}\alpha^\prime (\pm q_\ell +i p_\ell)~.
\end{align}
This identifies the associated 5-brane junction. Explicit examples will be discussed below.

\subsection{Conformal defects}\label{sec:conf-hol}

For conformal defects the D3-branes are localized at a point $w_c$ on $\Sigma$ and on the $S^2$, while wrapping an $\rm AdS_4$ subspace of $\rm AdS_6$ at that point. The embedding on $\Sigma$ does not depend on the choice of $\rm AdS_6$ coordinates, so we can uniformly describe planar defects in $\RR^5$, $\rm AdS_3$ defects in the field theory on $\rm AdS_5$, and $S^3$ defects in $S^5$. This will be different for defect RG flows.
As shown in \cite{Gutperle:2020rty}, the location $w_c$ on $\Sigma$ is determined by 
\begin{align}\label{eq:conf-defect-BPS}
	\partial_w\mathcal G\big\vert_{w=w_c}&=0~.
\end{align}
For regular $\rm AdS_6$ solutions and D3-branes in the interior of $\Sigma$ this condition implies
\begin{align}
	\cA_+(w)-\overline{\cA}_-(w)&=0~.
\end{align}
For regular 5-brane web solutions $\cG$ has a unique extremum in $\Sigma$.\footnote{For solutions obtained by T-duality from the Brandhuber/Oz solution in Type IIA \cite{Lozano:2012au,Lozano:2013oma}, the functions $\cA_\pm$ were given in \cite{DHoker:2016ujz,Hong:2018amk,Lozano:2018pcp}. For these solutions $\cG$ has no extremal points in $\Sigma$. Defects in Type IIA were discussed in \cite{Penin:2019jlf}.}
In the following we discuss the supergravity solutions associated with the brane webs in fig.~\ref{fig:pqwebs} and describe conformal 3d surface defects holographically, before moving on to defect RG flows. The D3-brane action for the conformal embeddings is given by \cite[(3.10)]{Gutperle:2020rty}
\begin{align}
	S_{\rm D3}&=6 T_{\rm D3}\Vol_{\rm AdS_4}\cG\big\vert_{w=w_c}~.
\end{align}
For global $\rm AdS_4$, corresponding to a conformal $S^3$ defect, the renormalized volume is given by $\Vol_{\rm AdS_4}=\frac{2}{3}\Vol_{S^3}$, with $\Vol_{S^3}=2\pi^2$, and $T_{\rm D3}^{-1}= (2 \pi)^3 (\alpha ^{\prime})^2$. In the following we set $2\pi\alpha^\prime=1$. The function $\cG$ vanishes on $\partial\Sigma$ and is positive in the interior, so  (\ref{eq:conf-defect-BPS}) has a solution.

\bigskip
\textbf{\boldmath{$+_{N,M}$} theory:} 
For the solutions associated with the brane webs in fig.~\ref{fig:pqwebs}, $\Sigma$ can be taken as the upper half plane with complex coordinate $w$.
The functions $\cA_\pm$ and $\cG$ were given explicitly in \cite{Uhlemann:2020bek}. 
For the $+_{N,M}$ web (after a coordinate transformation),
\begin{align}\label{eq:cA-plus-N}
	\cA_\pm&=\frac{3}{8\pi}\left[-iN\ln w\pm M\ln \frac{ w+1}{ w-1}\right]~, 
	&
	\cG&=\frac{9NM}{8\pi^2}\left[D\left(\frac{w+1}{ w-1}\right)+D\left(\frac{1-w}{1+w}\right)\right]~,
\end{align}
where $D$ is the Bloch--Wigner dilogarithm function \cite{Zagier:2007knq}
\begin{equation}\label{eq:BWdilog}
	D(u) = \Im \left[ \Li_2 (u) + \ln \lvert u \rvert \log (1-u) \right] .
\end{equation}
The external 5-branes emerge at the poles of $\partial_w\cA_\pm$ at $w\in\lbrace 0,\pm 1,\infty\rbrace$.
The conformal D3-brane defect location and action are (taking global $\rm AdS_4$ to describe an $S^3$ defect in $S^5$)
\begin{align}
	w_c&=i~,&  S_{\rm D3}&= \frac{9C}{\pi} MN ,
\end{align}
where $C$ is Catalan's constant. This gives the defect contribution to the free energy. It is quadratic in the 5-brane charges, compared to the quartic scaling of the free energy of the ambient 5d SCFT.
The $+_{N,M}$ solution has a $\ZZ_2\times \ZZ_2$ symmetry, corresponding to horizontal and vertical reflections in the brane web in fig.~\ref{fig:pqwebs}. For $N=M$ it has a $\ZZ_4$ symmetry under a combination of $SL(2,\RR)$ transformation of the upper half plane and Type IIB $\rm SL(2,\RR)$ transformations, discussed in \cite{Kim:2021fxx,Apruzzi:2022nax}. The D3-brane is located at the fixed point of these symmetries.

\bigskip
\textbf{\boldmath{$T_N$} theory:} 
For the $T_N$ solutions $\Sigma$ is again the upper half plane, but now $\partial_w\cA_\pm$ have 3 poles on the boundary, encoding the 3 groups of external 5-branes in fig.~\ref{fig:pqwebs}:
\begin{align}\label{eq:cApmTN}
	\cA_\pm^{T_N}&=\frac{3N}{8\pi} \left[\pm \ln\left(\frac{w-1}{w+1}\right)+i\ln\left(\frac{2w}{w+1}\right)\right]~, \nonumber \\
	\cG_{T_N}&=\frac{9N^2}{8\pi^2}D\left(\frac{2w}{w+1}\right)~.
\end{align}
The poles on $\partial\Sigma$ at $w\in\lbrace 0,\pm 1\rbrace$ encode the charges of the 5-brane junction via (\ref{eq:charges}).
The conformal D3-brane location and action are
\begin{align}
	w_c&=\frac{i}{\sqrt{3}}~,
	&
	S_{\rm D3}&= \frac{9}{2\pi} N^2 \Im \Li_2 \left( e^{i \pi /3} \right) .
\end{align}
The $T_N$ theory and the associated solution have a $\ZZ_3$ symmetry, discussed in \cite{Acharya:2021jsp,Kim:2021fxx,Apruzzi:2022nax}, and, as in the $+_{N,M}$ case before, the conformal defect D3-brane is located at the fixed point of this symmetry.

\bigskip
\textbf{\boldmath{$Y_N$} theory:} 
The $Y_N$ solution, corresponding to the junction of $N$ $(1,1)$ and $N$ $(-1,1)$ 5-branes with $2N$ NS5-branes in fig.~\ref{fig:pqwebs}, is specified by
\begin{align}\label{eq:YN-cApm}
	\cA^{Y_N}_\pm&=\frac{3N}{8\pi} \left[i\ln\left(\frac{w-1}{w+1}\right)\pm\ln\left(\frac{w^2-1}{4w^2}\right)\right]~, \nonumber \\
	\cG_{Y_{N}}&=\frac{9N^2}{4\pi^2}D\left(\frac{2w}{w+1}\right)~.
\end{align}
The conformal defect D3-brane is specified by 
\begin{align}\label{eq:YN-conf-SD3}
	w_c&=\frac{i}{\sqrt{3}}~,
	&
	S_{\rm D3}&= \frac{9}{\pi} N^2 \Im \Li_2 \left( e^{i \pi /3} \right) .
\end{align}
The $Y_N$ solution has a $\ZZ_3$ symmetry analogous to the $T_N$ solution, and the D3-brane is again located at the fixed point.
Moreover, the $Y_N$ solution is related by a Type IIB $SL(2,\RR)$ transformation combined with a rescaling of $N$ to the $T_N$ solution. The field theories engineered by the corresponding 5-brane junctions are different, but certain quantities are related at large $N$.

\subsection{Defect RG flows: BPS conditions}\label{sec:BPS-flow}

For defect RG flows, the choice of conformal representative of the field theory geometry matters. The conformal group of the 5d SCFT is $SO(2,5)$, and a conformal 3d defect preserves an $SO(2,3)$ subgroup. Defect RG flows on a planar defect extending along $\RR^{1,2}$ in a 5d SCFT on $\RR^{1,4}$ preserve the $ISO(1,2)$ isometries of $\RR^{1,2}$. Such defect RG flows were discussed holographically in \cite{Gutperle:2020rty}. 

Our main interest is in defects extending along an $S^3$ in $S^5$, which preserve $SO(4)$. This allows for comparison to localization calculations. 
The curvature of the spheres introduces an additional scale, and the BPS equations become more involved. Moreover, for non-conformal QFTs on spheres one can realize either supersymmetry or reflection positivity, but not both (similar to the tension between unitarity and supersymmetry on de Sitter space). As a result, in the holographic duals for non-conformal QFTs on $S^5$ (or defects on $S^3$) the usual reality conditions have to be relaxed.\footnote{%
In field theory terms, relevant deformations need to be accompanied by curvature couplings which spoil reality of the action in order to preserve supersymmetry. These additional couplings in turn impact the holographic duals.}
To solve these complications one at a time, we start by reviewing planar defect RG flows, then discuss RG flows for $\rm AdS_3$ defects in 5d SCFTs on $\rm AdS_5$, and then obtain the desired defect RG flows for $S^3$ defects in 5d SCFTs on $S^5$ by analytic continuation. 
The technical derivations are relegated to appendix~\ref{app:BPS}; here we only summarize the results.

For the defect RG flows we consider the D3-brane remains at a fixed point on the $S^2$ in ${\rm AdS}_6\times S^2\times \Sigma$, so that a $\mathfrak{u}(1)$ symmetry of the $S^2$ isometries is preserved. The flows also preserve the $\mathfrak{u}(1)$ symmetry corresponding to rotations transverse to the defect in the ambient CFT geometry.
	
\bigskip

\textbf{Planar defects:}
For a 5d SCFT on $\RR^{1,4}$ we choose as $\rm AdS_6$ geometry
\begin{align}\label{eq:AdS6-metric-flat}
	ds^2_{\rm AdS_6}&=dr^2+e^{2r}ds^2_{\RR^{1,4}}~.
\end{align}
The embedding for conformal D3-brane defects amounts to wrapping an $\RR^{1,2}$ subspace of $\RR^{1,4}$ at a fixed location on $\Sigma$. For defect RG flows the embedding is generalized to allow the location on $\Sigma$ to depend on the $\rm AdS_6$ radial coordinate. The position on $S^2$ is kept fixed so as to preserve a $\mathfrak{u}(1)$ subalgebra of the R-symmetry. 
The BPS condition was derived in  \cite{Gutperle:2020rty} and reads
\begin{equation}\label{eq:BPS-D3-flat}
	\cA_{+} (w) - \overline{\cA}_{-} (\bar w) = m e^{-r}~,
\end{equation}
where $m$ is a complex parameter and $w (r)$ is the BPS embedding solution.
The conformal embedding is recovered for $m=0$.
Turning on $m$ triggers operators with $\Delta=2$ in standard quantization or $\Delta=1$ in alternative quantization.
The embeddings for a single D3-brane start at the point on $\Sigma$ where the conformal D3-brane would be at the conformal boundary of $\rm AdS_6$ ($r=\infty$) and approach one of the poles on the boundary of $\Sigma$ as $r$ decreases.\footnote{For a pair of D3-branes ending on the brane web from opposite sides, more general mass deformations are possible. Then D3-brane embeddings can end smoothly on generic boundary points of $\Sigma$ in the IR. For details see \cite{Gutperle:2020rty}.}
The action evaluated on BPS embeddings is derived in app.~\ref{app:BPS} and given by 
\begin{align}
	S_{\rm D3} &=2T_{\rm D3} \Vol_{\RR^{1,2}} \int dr\, \frac{d \ }{dr}\left[e^{3r}\cG\right]~.
\end{align}
That is, the action is given by $\cG$ evaluated at the end points of the embeddings. The endpoint at $r\rightarrow\infty$ has to be treated in the usual way with holographic renormalization.

\bigskip

\textbf{Defects on \boldmath{$\rm AdS_3\subset AdS_5$}:}
We now place the 5d SCFT on two copies of Poincar\'e $\rm AdS_5$ joined at their conformal boundaries. For the 5d SCFT itself, this is just a different description of $\RR^{1,4}$, as each $\rm AdS_5$ is conformal to half of $\RR^{1,4}$.
But we add a defect along an $\rm AdS_3$ subspace and study defect RG flows. This choice of geometry leads to a different natural notion of mass deformation. Defect RG flows on $\rm AdS_3$ preserve the $SO(2,2)$ isometries of $\rm AdS_3$, which differs from planar defects.\footnote{A (constant) mass deformation in flat space corresponds to a position-dependent mass in AdS and vice versa.}
To describe the 5d SCFT on two copies of $\rm AdS_5$, we take $\rm AdS_6$ in $\rm AdS_5$ slicing,
\begin{align}\label{eq:metric-AdS6-AdS5}
	ds^2_{\rm AdS_6}&=dr^2+ \cosh^2\!r\,ds^2_{\rm AdS_5}~,
\end{align}
with the conformal boundary comprising the two regions $r\rightarrow \pm\infty$.
The D3-brane wraps an $\rm AdS_3$ inside $\rm AdS_5$.
The BPS condition is derived and integrated in app.~\ref{app:BPS}; the result is
\begin{align}\label{eq:BPS-AdS}
	(\sinh r-i)\cA_+(w)-(\sinh r+i)\overline{\cA}_-(\bar w)&=\frac{m}{2}~.
\end{align}
For large $|r|$ this approaches the defect condition for planar defects in (\ref{eq:BPS-D3-flat}), as expected.
The D3-brane action evaluated on BPS solutions is given by
\begin{align}
	S_{\rm D3} &= 2T_{\rm D3} \Vol_{\rm AdS_3} \int dr\,
	\frac{d}{dr}\left[h(r) \cG
	-\frac{1}{2}(\overline{m}\cA_++m\cA_-+m\overline{\cA}_++\overline{m}\overline{\cA}_-)+2i\cB-2i\overline{\cB}
	\right],
\end{align}
where $h(r)=\frac{1}{4}(\sinh(3r)+9\sinh(r))$ satisfies $h'(r)=3\cosh^3\!r$.
These results provide the basis for studying the phase structure of $\rm AdS_3$ defects in $\rm AdS_5$. This would be expected to give supersymmetric realizations of ``Janus on the brane" embeddings \cite{Gutperle:2020gez}.

\bigskip

\textbf{Defects on \boldmath{$S^3\subset S^5$}:}
Finally, we come to 5d SCFTs on $S^5$ with a 3d defect extending along an $S^3$.
The $\rm AdS_6$ metric describing the 5d SCFT on $S^5$ is given by
\begin{align}\label{eq:AdS6-metric-S5}
	ds^2_{\rm AdS_6}&=dr^2+ \sinh^2\!r\,ds^2_{S^5}~.
\end{align}
This metric is related to the one for $\rm AdS_6$ in $\rm AdS_5$ slicing in (\ref{eq:metric-AdS6-AdS5}) by the analytic continuation
\begin{align}\label{eq:S3-cont}
	r\rightarrow r+\frac{i\pi}{2}~.
\end{align}
The D3-brane wraps an $S^3$ in the $S^5$ in (\ref{eq:AdS6-metric-S5}) and the BPS condition can be obtained from the condition for $\rm AdS_3$ defects in (\ref{eq:BPS-AdS}) by implementing the analytic continuation (\ref{eq:S3-cont}).
Crucially, for the analytic continuation the condition in (\ref{eq:BPS-AdS}) and its complex conjugate have to be treated as independent equations.
The analytic continuation then leads to the two conditions
\begin{align}\label{eq:BPS-D3-S3}
	(\cosh r-1)\cA_+-(\cosh r+1)\overline{\cA}_-&=-\frac{im}{2}~,
	\nonumber\\
	(\cosh r+1)\overline{\cA}_+-(\cosh r-1)\cA_-&=-\frac{i\overline{m}}{2}~.
\end{align}
After the analytic continuation the two equations are not related by complex conjugation anymore. This is a common effect for supersymmetry in Euclidean signature; we regard $w$ and $\bar w$, and hence $\cA_\pm$ and $\overline{\cA}_\pm$, as independent quantities.
Likewise, $m$ and $\overline{m}$ are not necessarily related by complex conjugation in Euclidean signature.
The action evaluated on BPS embeddings becomes
\begin{align}\label{eq:D3-action-S3}
	S_{\rm D3} = 2T_{\rm D3} \Vol_{\rm S^3} \int dr\,
	\frac{d}{dr}\left[ \tilde{h}(r) \cG+2\cB-2\overline{\cB}
	-\frac{i}{2}\left(\overline{m}\cA_++m\cA_-+m\overline{\cA}_++\overline{m}\overline{\cA}_-\right)
	\right],
\end{align}
where $\tilde{h}(r)=\frac{1}{4}(\cosh(3r)-9\cosh(r))$ satisfies $\tilde{h}'(r)=3\sinh^3\!r$.
These results are the basis for the following calculations.

\subsection{Defect RG flows: case studies}\label{sec:AdScase}

We now use the BPS conditions for $S^3$ defects in 5d SCFTs on $S^5$ to discuss the D3-brane embeddings and defect contribution to the free energy. We will primarily focus on embeddings which reach from $r=\infty$ all the way to $r=0$, where the D3-brane can cap off smoothly since the $S^3$ in $\rm AdS_6$ collapses. For the calculation of the defect contribution to the free energy we focus, for simplicity, on the part with highest transcendental weight: the free energy, given by the D3-brane action in \eqref{eq:D3-action-S3}, generally comprises dilogarithm terms as well as logarithmic parts. Focusing on the dilogarithm terms, we can drop the contribution from the conformal boundary, including holographic counterterms, which on dimensional grounds are polynomials in $m, \overline{m}$. 
We leave such terms as ambiguous and focus on the unambiguous dilogarithm terms. This will be sufficient for a meaningful comparison to the localization computations in sec.~\ref{sec:loc-tot}.
The contribution from $r=0$ is
\begin{align}\label{eq:SD3-S}
	S_{\rm D3, IR} = 2T_{\rm D3} \Vol_{\rm S^3} 
	\left[2\cA_+\overline{\cA}_+ -2\cA_-\overline{\cA}_-+4\overline{\cB}
	+\frac{i}{2}\left(\overline{m}\cA_++m\cA_-+m\overline{\cA}_++\overline{m}\overline{\cA}_-\right)
	\right]_{r=0}.
\end{align}
The BPS equations (\ref{eq:BPS-D3-S3}) for $r=0$ yield
\begin{align}
	-2\overline{\cA}_-\big\vert_{r=0}&=-\frac{im}{2}~,
	&
	2\overline{\cA}_+\big\vert_{r=0}&=-\frac{i\overline{m}}{2}~.
\end{align}
In the action (\ref{eq:SD3-S}) the terms proportional to $\cA_\pm$ cancel, and we arrive at 
\begin{align}\label{eq:SD3-S-2}
	S_{\rm D3, IR} = 8T_{\rm D3} \Vol_{\rm S^3} 
	\overline{\cB}\big\vert_{r=0}~.
\end{align}
Up to the contribution from the conformal boundary, the on-shell action is determined by $\bar w$ alone.

\bigskip

\textbf{\boldmath{$+_{N,M}$} theory:} We start with the $+_{N,M}$ solution with $\cA_\pm$ and $\cG$ in (\ref{eq:cA-plus-N}) and the brane web in fig.~\ref{fig:pqwebs}.
The conformal defect is at $w_c=i$. 
For planar mass-deformed defects, embeddings extending all the way into the IR either proceed along the imaginary axis to one of the poles at $w=0,\infty$, or along a semi-circle connecting $w_c$ to $w=\pm 1$. These embeddings were discussed in \cite{Gutperle:2020rty}.

To study embeddings describing $S^3$ defects we introduce a new coordinate $z$ on $\Sigma$, such that
\begin{align}
	w&=e^z~, & \Sigma&=\left\lbrace z\in\CC\,\vert\,0\leq \Im(z)\leq \pi\right\rbrace~.
\end{align}
In the $z$ coordinate $\Sigma$ is an infinite strip, and
\begin{align}
	\cA_\pm&=\frac{3}{8\pi}\left[-iNz\pm M\ln \coth\left(\frac{z}{2}\right)\right].
\end{align}
The poles at $w=0,\pm1,\infty$ are mapped to $z=0,i\pi$ and $\Re(z)=\pm\infty$.
The $\ZZ_2$ symmetry axes corresponding to horizontal and vertical reflections in the brane web correspond to $\Re(z)=0$ and $\Im(z)=\frac{\pi}{2}$, respectively.
In the $z$ coordinate we have
\begin{align}
	\overline\cB&=
	\frac{9 iNM}{32 \pi ^2}
	\left[2 \Li_2\left(-e^{-\bar z}\right)-2 \Li_2\left(e^{-\bar z}\right)+\bar z\ln\tanh
	\left(\frac{\bar z}{2}\right)\right]~.
\end{align}

To rewrite the BPS conditions more conveniently, we further introduce 
\begin{align}
	h_{\pm}&\equiv \cA_+ \pm \cA_-~,
\end{align}
The BPS conditions \eqref{eq:BPS-D3-S3} become
\begin{align}\label{eqn:BPS-h}
	2(\cosh r-1)h_+ -2(\cosh r+1)\overline{h}_+&=-i(m-\overline{m})~,\nonumber\\
	2(\cosh r-1)h_- +2(\cosh r+1)\overline{h}_-&=-i(m+\overline{m})~.
\end{align}
The explicit expressions for the $+_{N,M}$ solution are
\begin{align}
	h_+&=\frac{3N}{4\pi}(-iz)~, & h_-&=\frac{3M}{4\pi}\ln\coth\left(\frac{z}{2}\right)~,
\end{align}
and we define $m=3(M m_1+i N m_2)/(2\pi)$ and $\overline{m}=3(M m_1-iNm_2)/(2\pi)$.
In Euclidean signature $m_{1/2}$ can be complex. 
The conditions in (\ref{eqn:BPS-h}) become
\begin{align}
	(\cosh r-1)z+(\cosh r+1)\overline{z}&=-2im_2~,
	\nonumber\\
	(\cosh r-1)\ln\coth\left(\frac{z}{2}\right)+(\cosh r+1)\ln\coth\left(\frac{\bar z}{2}\right)&=-2im_1~.
\end{align}
The conformal embedding corresponds to $m_1=m_2=-\frac{\pi}{2}$ and $z=\frac{i\pi}{2}=-\bar z$.
In general, $z$ and $\bar z$ are not necessarily related by analytic continuation.
The first equation can be solved for $\bar z$ in terms of $z$. The second equation then becomes an equation solely for $z$.
For embeddings reaching all the way to $r=0$, the BPS conditions have to be enforced at $r=0$.
From the $r=0$ equations we find
\begin{align}\label{eq:plus-m2-m1}
	\bar z\vert_{r=0} &=im_2~,
	&i\ln\coth \left(\frac{im_2}{2}\right)&=m_1~.
\end{align}
This can be used to eliminate, say, $m_2$ in favor of $m_1$. The result is one complex equation for $z$ with one complex parameter, which is either $m_1$ or  $m_2$.

Thanks to the integrated form of the D3-brane action in (\ref{eq:D3-action-S3}), we do not need the entire embedding explicitly to evaluate the defect free energy.
We define 
\begin{align}\label{eq:m1-mu-plus}
	m_1&=-\frac{\pi}{2}-2\pi i\mu~,
\end{align}
such that $\mu=0$ corresponds to the conformal defect.
As explained above, we focus on the contribution from $r=0$ in (\ref{eq:SD3-S-2}) for simplicity; with $\bar z\vert_{r=0}$ in (\ref{eq:plus-m2-m1}) we find
\begin{align}\label{eq:SD3-plus}
	\Re S_{\rm D3, IR}&=
	\frac{9 M N}{4 \pi } \Im \left(2\Li_2\left(i e^{-2 \pi\mu}\right)-2\Li_2\left(-i e^{-2 \pi\mu}\right)+m_1m_2\right)
	~.
\end{align}
Properly accounting for the UV contribution to the action from $r=\infty$ may contribute additional logarithmic terms, which may arise from polynomials in $m_1$, $m_2$ due to the relation in (\ref{eq:plus-m2-m1}), and further scheme-dependent counterterms may be added, which are at most polynomial in $\mu$.
The dilogarithm terms in (\ref{eq:SD3-plus}), however, are unambiguous.
They will be compared to (\ref{massiveFDpmn}) below.

\begin{figure}
	\includegraphics[height=0.2\linewidth]{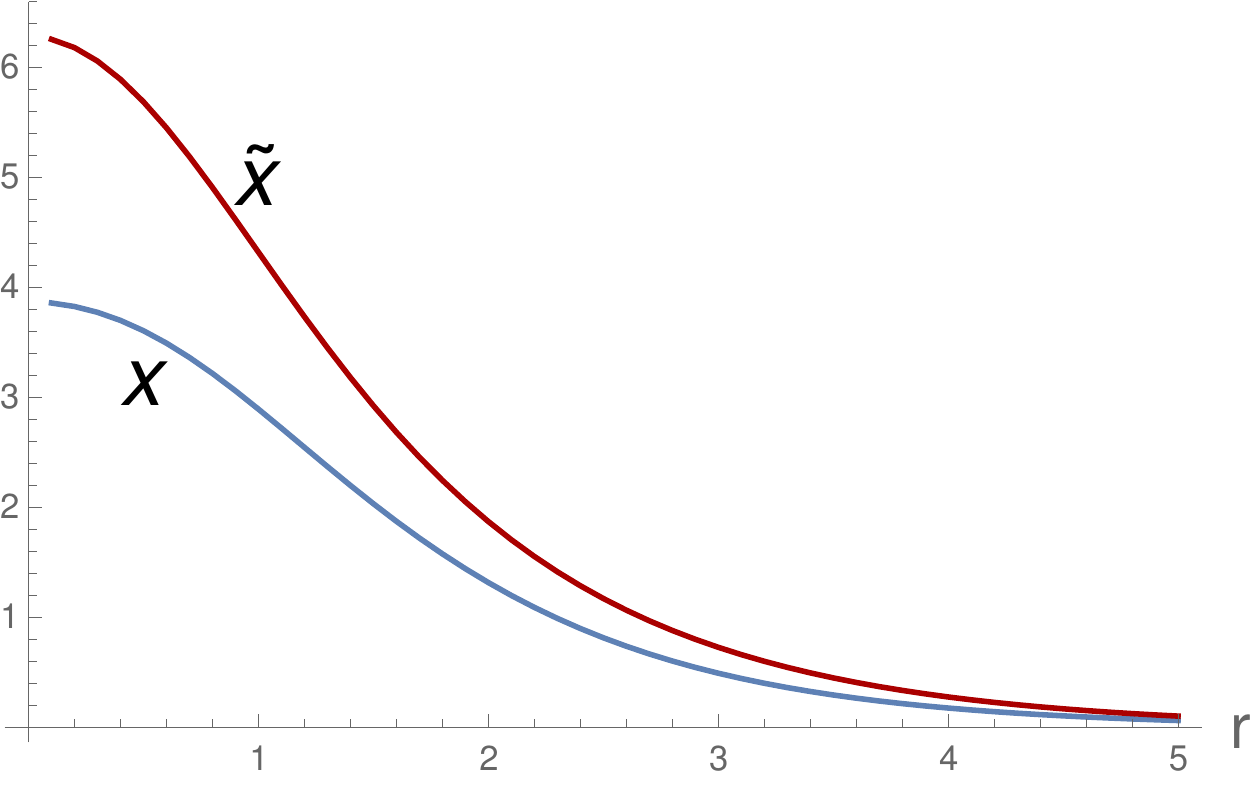}
	\hskip 10mm
	\includegraphics[height=0.2\linewidth]{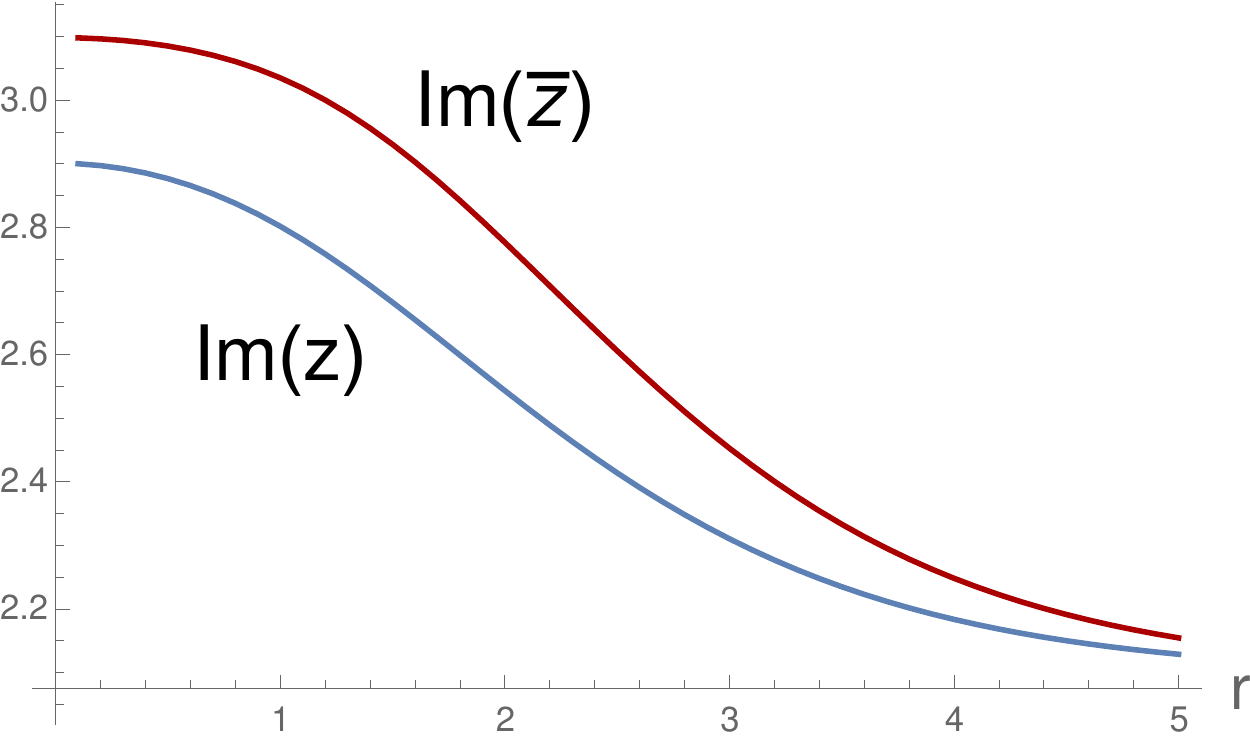}
	\caption{Left: D3-brane embeddings for the $+_{N,M}$ solutions along the $\ZZ_2$ fixed line $\Im(z)=\frac{\pi}{2}$, as in (\ref{eq:plus-ansatz}), for $\mu=-1$. Right: D3-brane embeddings along the imaginary axis for the $Y_N$ solution \eqref{eq:YN-cA-z} with $4\pi\mu=1$.}
	\label{fig:S3-plot}
\end{figure}

The $+_{N,M}$ junction in fig.~\ref{fig:pqwebs} has a $\ZZ_2\times \ZZ_2$ symmetry corresponding to horizontal and vertical reflections. One $\ZZ_2$ symmetry is represented in the quiver gauge theory (\ref{eq:D5NS5-quiver}) as reflection across the central gauge node.
We therefore discuss D3-brane embeddings which are consistent with one of the $\ZZ_2$ symmetries of the solution more explicitly. In the brane junction in fig.~\ref{fig:pqwebs} they correspond to a horizontal or vertical displacement of the D3-brane, i.e.\ a displacement along the fixed line of the horizontal or vertical reflection symmetry. Embeddings along the two fixed lines are related by S-duality, so it suffices to discuss one of them.
We start with the ansatz
\begin{align}\label{eq:plus-ansatz}
	z&=x+\frac{i\pi}{2}~, & \bar z&=\tilde x+\frac{i\pi}{2}~,
\end{align}
where $x$, $\tilde x$ are real. Solutions can then be found if $\mu$ in (\ref{eq:m1-mu-plus}) is real as well. The solutions obtained from this ansatz connect the conformal point $x=0$ to a cap-off point on the line $\Im(z)=\frac{\pi}{2}$. The embedding is determined by an algebraic equation, as detailed above. An example plot is shown in fig.~\ref{fig:S3-plot}.
The plots show that $x'(0)=\tilde x'(0)=0$ and thus $z'(0)=\bar z'(0)=0$; this ensures that the D3-brane caps off at $r=0$ smoothly with the $S^3$ inside $\rm AdS_6$ collapsing in such a way that the induced D3-brane metric has no conical singularity.

\bigskip

\textbf{\boldmath{$T_N$} theory:} For the $T_N$ theory we follow a similar approach as for the $+_{N,M}$ theories: we use a coordinate transformation to make one of the two BPS equations straightforward to solve and then leave the remaining equation to determine the embedding. For the $T_N$ theory we use
\begin{align}\label{eq:TN-w-z}
	\frac{w-1}{w+1}&=e^z~.
\end{align}
As for the $+_{N,M}$ theory, the Riemann surface $\Sigma$ is a strip $0<\Im(z)<\pi$ in this coordinate. We have
\begin{align}
	\cA_\pm&=\frac{3N}{8 \pi }\left(\pm z+i \ln \left(1+e^{-z}\right)\right)~,
	&
	\overline{\cB}&=\frac{9 i N^2}{32 \pi ^2}\left(2 \text{Li}_2\left(-e^{-\bar z}\right)-\bar z\ln \left(1+e^{-\bar z}\right)\right)~.
\end{align}
The poles at $w=0,\pm 1$ are mapped to $\Re(z)=\pm\infty$ and $z=i\pi$.
The BPS equations (\ref{eq:BPS-D3-S3}) become
\begin{align}
	\cosh r\, \ln \left((1+e^{-z})(1+e^{-\bar z})\right)-\ln \left(\frac{1+e^{-z}}{1+e^{-\bar z}}\right)&=-2im_2~,
	\nonumber\\
	\cosh r (z+\bar z)-z+\bar z&=-2im_1~,
\end{align}
where $2\pi m=3N(m_1+im_2)$ and $2\pi\overline{m}=3N(m_1-im_2)$. 
The conformal defect corresponds to 
\begin{align}
	m_1&=\frac{2\pi}{3}~, &m_2&=-\frac{\pi}{3}~,&  z&=\frac{2\pi i}{3}~.
\end{align}
We will again focus on D3-brane embeddings which extend from $r=\infty$ all the way to $r=0$ where they cap off smoothly. 
From the $r=0$ equations we find
\begin{align}
	\ln\left(1+e^{-\bar z}\right)\vert_{r=0}&=-im_2~, & \bar z\vert_{r=0}&=-i m_1~.
\end{align}
We can thus eliminate $m_2$ in favor of $m_1$. 
With 
\begin{align}
m_1&=2\pi/3+2\pi i \mu~,
\end{align}
such that $\mu=0$ corresponds to the conformal defect, we find
\begin{align}\label{eq:SD3-TN-IR}
	\Re S_{\rm D3,IR}&=\frac{9N^2}{4\pi}\Im\left(2\Li_2\left(e^{-2\pi\mu+i\pi/3}\right)-m_1m_2\right)~.
\end{align}
As for the $+_{N,M}$ result, the dilogarithm terms are unambiguous.
They will be compared to (\ref{massiveFDtn}).

The $\ZZ_3$ symmetry of the $T_N$ junction in fig.~\ref{fig:pqwebs} under rotations by $2\pi/3$ combined with an $SL(2,\ZZ)$ transformation is not manifest in the quiver gauge theory in (\ref{TNquiver}). We therefore have no symmetry-based argument to link the conformal defect to a particular gauge node, or single out deformations which would be distinguished from the gauge theory perspective. We will nevertheless identify a gauge node for the conformal defect from the supergravity perspective in sec.~\ref{sec:gauge-theory-connecttion}.

\bigskip
\textbf{\boldmath{$Y_N$} theory:} 
For the $Y_N$ theory, we start from the functions $\cA_\pm$ in \eqref{eq:YN-cApm} and again use the coordinate transformation (\ref{eq:TN-w-z}).
This leads to
\begin{align}
	\label{eq:YN-cA-z}
	\cA_\pm &=\frac{3N}{8\pi}\left((i\pm1)z\mp2\ln\left(1+e^z\right)\right)~,
	\nonumber\\
	\overline{\cB}&=
	\frac{9 i N^2 }{32 \pi ^2}
	\left(2 \Li_2\left(-e^{-\bar z}\right)-2 \Li_2\left(-e^{\bar z}\right)-\bar z \left(\bar z+2 \ln \left(1+e^{-\bar z}\right)\right)\right)~.
\end{align}
The poles are mapped to $\Re(z)=\pm\infty$ and $z=i\pi$. The solution has a $\ZZ_2$ symmetry under reflections across the line $\Re(z)=0$, corresponding to the horizontal reflection symmetry in the brane web in fig.~\ref{fig:pqwebs}.
The BPS equations (\ref{eq:BPS-D3-S3}) become
\begin{align}
	\cosh r (z+\bar z)-z+\bar z&=-2im_2~,
	\nonumber\\
	-\cosh r \ln \left((1+e^z)(1+e^{\bar z})\right)+ \ln
	\left(\frac{1+e^z}{1+e^{\bar z}}\right)&=-i (m_1-m_2)~.
\end{align}
The conformal defect corresponds to the embedding
\begin{align}\label{eq:YN-D3-conf-z}
	z&=\frac{2\pi i}{3}=-\bar z~, & m_1&=0~, &m_2&=\frac{2\pi}{3}~.
\end{align}
From the $r=0$ equations,
\begin{align}\label{eq:zbarBPSYN}
	\bar z\vert_{r=0}&=-im_2~, & 2\ln\left(1+e^{\bar z}\right)\vert_{r=0}&= i(m_1-m_2)~.
\end{align}
The D3-brane action becomes
\begin{align}
	\Re S_{\rm D3}&=\frac{9N^2}{4\pi}\Im \left[ 2\Li_2\left(-e^{- i m_2}\right)-2\Li_2\left(e^{- i m_2}\right)+m_1m_2\right]~.
\end{align}

Even though the family of $Y_N$ solutions is related to the $T_N$ family by $SL(2,\RR)$, and D3-brane embeddings can consequently be related between the solutions, the $Y_N$ and $T_N$ gauge theories are genuinely different, and so is the field theory interpretation of deformations displacing the D3-brane from the 5-brane junction in a particular direction -- say horizontally or vertically.

For the $Y_N$ theory the $\ZZ_2$ symmetry under horizontal reflection of the junction in fig.~\ref{fig:pqwebs} is represented in the quiver (\ref{eq:YN-quiver}) as reflection across the central node. 
We therefore discuss D3-brane embeddings which respect the $\ZZ_2$ symmetry of the $Y_N$ solution in detail.
We seek embeddings with $z$ and $\bar z$ both along the $\ZZ_2$ fixed lines $\Re(z)=\Re(\bar z)=0$ mentioned above, recalling that  $z(r)$, $\bar z (r)$ are independent embedding functions in Euclidean signature.
Such embeddings would naturally correspond to vertical displacements in the brane junction in fig.~\ref{fig:pqwebs}. 
Embeddings with $z$ and $\bar z$ along the imaginary axis need $m_1$ imaginary and $m_2$ real (as a necessary condition this can be derived from a near-boundary analysis).
We therefore use (\ref{eq:zbarBPSYN}) to eliminate $m_2$ in favor of $m_1$ and set 
\begin{align}\label{eq:YN-m1-mu}
	m_1&=4\pi i \mu~,
\end{align}
with $\mu$ real. An example solution is shown in fig.~\ref{fig:S3-plot}. Again, the plot shows that $z'(0)=\bar z'(0)=0$ such that the D3-brane caps off smoothly at $r=0$.
Solutions to the $r=0$ equations (\ref{eq:zbarBPSYN}) with imaginary $\bar z \vert_{r=0}$ can actually only be found if\footnote{For planar defects, whether a D3 flowing along the $\ZZ_2$ fixed line on $\Sigma$ can cap off smoothly in the IR is determined by the sign of the mass deformation \cite{Gutperle:2020rty}. On $S^3$ the additional scale allows for more general phase structure.}
\begin{align}\label{eq:YN-mu-crit}
	\mu&>\mu_\star~, & \mu_\star&=-\frac{\ln 2}{2\pi}~.
\end{align}
When $\mu$ approaches $\mu_\star$ from above, $\bar z\vert_{r=0}$ approaches zero along the positive imaginary axis.
For $\mu<\mu_\star$ the $r=0$ equation (\ref{eq:zbarBPSYN}) leads to $\bar z\vert_{r=0}$ on the real line away from the origin. The D3-brane flows then break the $\ZZ_2$ symmetry under reflection across the imaginary axis.
While $z(r)$ still proceeds along the imaginary axis, $\bar z$ connects the location of the conformal embedding in (\ref{eq:YN-D3-conf-z}) for $r=\infty$ to a point on the boundary of $\Sigma$ for $r=0$.
The D3-brane action expressed in terms of $\mu$ for $\mu>\mu_\star$ reads 
\begin{align}\label{eq:massiveSD3YNsugra}
	\Re S_{\rm D3}&=\frac{9N^2}{\pi}\Im \left[\Li_2\left(1+\frac{\sqrt{1-4e^{4\pi\mu}}-1}{2e^{4\pi\mu}}\right)+\pi \mu \ln \left(1+\frac{\sqrt{1-4e^{4\pi\mu}}-1}{2e^{4\pi\mu}}\right)\right].
\end{align}
For $\mu=0$ this result matches \eqref{eq:YN-conf-SD3}. The sign change in the arguments of the square roots reflects the transition in (\ref{eq:YN-mu-crit}).

In summary, for the $Y_N$ theory we find a phase transition in the defect free energy for mass deformations (\ref{eq:YN-m1-mu}) with real $\mu$: for $\mu>\mu_\star$ the defect RG flows triggered by a vertical displacement of the D3-brane preserve the $\ZZ_2$ symmetry under horizontal reflection, while for $\mu<\mu_\star$ the $\ZZ_2$ symmetry is broken.
This will be discussed from the gauge theory perspective in sec.~\ref{sec:M1FE}.

\subsection{5d/3d F-maximization in Type IIB}

The above discussion used the explicit supergravity duals for 5d SCFTs to obtain the defect free energy. However, we can extract a prescription for calculating the defect free energy for conformal defects which uses only the 5-brane charges as input. It takes the form of an extremization principle.

\begin{figure}
	\begin{tikzpicture}
		\foreach \i in {-1,0,1}{\foreach \j in {100,-50}{
				\draw[very thick] ({sin(\j)*\i*0.1},{-cos(\j)*\i*0.1}) -- ({cos(\j)+sin(\j)*\i*0.1},{sin(\j)-cos(\j)*\i*0.1}); 
		}}
		
		\foreach \i in {-3/4,3/4}{\foreach \j in {30,-110,-190}{
				\draw[very thick] ({sin(\j)*\i*0.1},{-cos(\j)*\i*0.1}) -- ({cos(\j)+sin(\j)*\i*0.1},{sin(\j)-cos(\j)*\i*0.1}); 
		}}
		\draw[fill=gray] (0,0) circle (5pt);
		\node[anchor=south] at ({cos(100)},{sin(100)}) {\scriptsize $(p_1,q_1)$};
		\node [anchor=south west] at ({cos(30)},{sin(30)})  {\scriptsize  $(p_2,q_2)$};
		\node [anchor=south west] at ({cos(-50)},{sin(-50)}) {\scriptsize  $(p_3,q_3)$};
		\node [rotate=125] at ({cos(-150)},{sin(-150)}) {\scriptsize  $\dots$};
	\end{tikzpicture}
	\caption{General $(p,q)$ 5-brane junction involving $L$ groups of 5-branes with charges $(p_1,q_1),\ldots ,(p_L,q_L)$.\label{fig:junction}}
\end{figure}
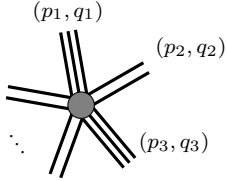

To this end, we start from the 5d $F$-maximization of \cite{Fluder:2020pym}: Consider a 5-brane junction involving $L$ groups of 5-branes with charges $(p_\ell,q_\ell)$, $\ell=1,...,L$ (fig.~\ref{fig:junction}) where the 5-brane charges are assumed to be large, of $\mathcal O(N)$. The free energy of the 5d SCFT on $S^5$ is obtained by introducing a real parameter $r_\ell$ for each 5-brane group, and solving the following extremization principle:
\begin{align}\label{eq:5d-extr}
	F_{S^5}&=\max_{\vec{r}\in \RR^L}F_{\rm trial}(\vec{r})~,
	\nonumber\\
	F_{\rm trial}(\vec{r})&=-\frac{9}{16\pi^2}\! \sum_{\ell,k,m,n=1}^L \!
	p^{}_{[\ell}q^{}_{k]} \cdot p^{}_{[m}q^{}_{n]}\cdot\mathcal L_3\!\left(\frac{r_k-r_m}{r_k-r_\ell}\frac{r_\ell-r_n}{r_m-r_n}\right),
\end{align}
where $p^{}_{[\ell}q^{}_{k]}\equiv p_\ell q_k -p_k q_\ell$ and $\mathcal L_3$ is the single-valued trilogarithm function
\begin{align*}
	\mathcal L_3(z)& =\Re\left[\Li_3(z)-\ln |z| \cdot \Li_2(z)-\frac{1}{3}\ln^2\!|z|\cdot \ln(1-z)\right]~.
\end{align*}
The sign convention is such that the 5d free energy is positive. 
In the extremization three of the $r_\ell$ can be fixed arbitrarily and the remaining ones are then determined by the extremization.\footnote{In terms of the supergravity duals $r_\ell$ are the locations of poles on $\partial\Sigma$; the freedom to choose three $r_\ell$ corresponds to $SL(2,\RR)$ transformations of $\Sigma$.}
It was shown in \cite{Fluder:2020pym} that this generally yields the free energy on $S^5$. 
Both extremality ($\partial F(\vec{r})/\partial r_\ell=0$) and maximality ($\partial^2 F(\vec{r})/(\partial r_\ell \partial r_k)$ negative definite once three of the $r_\ell$ are fixed) are needed.

This prescription can be extended to conformal 3d defects described by D3-branes ending on the 5-brane junction: Based on the general results of \cite{Gutperle:2020rty}, the defect free energy for a conformal defect is given by
\begin{align}
	F_{S^3\subset S^5}&=\max_{w\in \mathds{H}} F_{\rm trial, 3d}(w)~,
	&
	F_{\rm trial,3d}(w)&=\frac{9}{4\pi} \sum_{\stackrel{\ell,k=1}{\ell\neq k}}^L p^{}_{[\ell}q^{}_{k]}D\left(\frac{r_\ell-w}{r_\ell-r_k}\right),
\end{align}
where $\mathds{H}$ is the upper half complex plane and $D$ is the Bloch--Wigner dilogarithm function \eqref{eq:BWdilog}.
The parameters $r_\ell$ are determined by the 5d $F$-maximization (\ref{eq:5d-extr}).
The defect free energy is $\mathcal O(N^2)$, compared to the 5d free energy of $\mathcal O(N^4)$.

We will connect the holographic results for defects in 5d SCFTs to gauge theory descriptions of the 5d theories and 3d defects. This will lead to an independent $F$-maximization principle on the field theory side (sec.~\ref{sec:CFE}). It would be interesting to understand if these procedures are related.

\subsection{Connection to gauge theories} \label{sec:gauge-theory-connecttion}
The supergravity description provides an interesting further piece of information which will be relevant below: By studying Wilson loop operators $\cW_\wedge$ in antisymmetric representations of individual gauge nodes, which are represented holographically by D3-branes localized at a point on $\Sigma$ and wrapping the $S^2$ as well as an $\rm AdS_2$ in $\rm AdS_6$, one can define a coordinate system on $\Sigma$ which has a precise meaning in the 5-brane web associated with a given supergravity solution \cite{Uhlemann:2020bek}. Namely, one can introduce  discrete coordinates on the brane web labeling the open faces in the resolved web, e.g.\ for the resolved $+_{N,M}$ web in fig.~\ref{fig:pqwebs} running from $1$ to $M-1$ horizontally and from $1$ to $N-1$ vertically.
Then, upon defining
\begin{align}\label{eq:NF1-ND1-cApm}
	N_{\rm F1}+i N_{\rm D1}&=\frac{4}{3}\left(\cA_+ +\overline{\cA}_-\right)\,,
\end{align}
$N_{\rm D1}$ corresponds to the horizontal coordinate on the brane web and $N_{\rm F1}$ to the vertical coordinate. For a given point on $\Sigma$, the horizontal coordinate identifies the gauge node which a Wilson loop D3-brane at this point is associated with. 

We can use the coordinate in (\ref{eq:NF1-ND1-cApm}) to also link the conformal 3d defect D3-branes to particular gauge nodes, labeled by $t_0$ in the notation of fig.~\ref{fig:quiver}. 
Gauge theory descriptions for the $+_{N,M}$, $T_N$ and $Y_N$ theories are given below in (\ref{eq:D5NS5-quiver}), (\ref{TNquiver}) and (\ref{eq:YN-quiver}), respectively.
The identification in (\ref{eq:NF1-ND1-cApm}) links the conformal 3d defects to the following gauge nodes:
\begin{align}\label{eq:gauge-nodes-sugra}
	+_{N,M}:\quad t_0&=\frac{1}{2}M~,
	&
	T_N:\quad t_0&=\frac{2}{3}N~,
	&
	Y_N:\quad t_0&= N~.
\end{align}
For the $+_{N,M}$ and $Y_N$ theories this identification can also be argued for based on the $\ZZ_2$ symmetries: the D3-brane is located at the fixed point of a $\ZZ_2$ symmetry acting as reflection of the quiver across the central node. But for the $T_N$ theory there is no such simple argument.
This identification will be further supported in sec.~\ref{sec:CFE} below.

In \cite{Legramandi:2021uds,Fatemiabhari:2022kpv}, Type IIB $\rm AdS_6$ solutions dual to 5d SCFTs with gauge theory deformations as balanced quivers were formulated in a way which directly connects to gauge theory language. This covers the $+_{N,M}$ and $T_N$ solutions, and the S-dual of the $Y_N$ solution has a balanced quiver description given in (\ref{eq:YN-quiver-S}). It would be interesting to study the D3-brane defects from that perspective as well (including their fate upon compactification as in \cite{Legramandi:2021aqv}).

We close this part with an empirical observation: The expectation value for a Wilson loop in the antisymmetric representation associated with a particular gauge node is given by  \cite{Uhlemann:2020bek}
\begin{align}
	\langle \cW_\wedge\rangle& = \frac{8 \pi}{3} \cG \left(w\right),
\end{align}
where $w$ denotes the position of the corresponding D3-brane on $\Sigma$. Through eq.~(\ref{eq:NF1-ND1-cApm}), $w$ identifies the gauge node and rank of the representation of the Wilson loop. Comparing to the conformal codimension-2 operators, the two are governed by the same function $\cG$, and the BPS condition in (\ref{eq:conf-defect-BPS}) means that the D3-brane representing a conformal 3d defect is located at the same point on $\Sigma$ as the D3-brane representing the Wilson loop with overall maximal expectation value.

\section{Localization and defect free energy}\label{sec:loc-tot}

In this section we discuss 3d defects in 5d gauge theories using supersymmetric localization, and connect the results to those of the previous section for 5d SCFTs. The matrix models will be constructed by combining general building blocks for the matrix models resulting from localization in separate 3d and 5d theories. A formal justification and derivation of the one-loop determinant in the matrix model for defect multiplets on $S^3$ is given in sec.~\ref{sec:localization}. In sec.~\ref{sec:localization-F} and \ref{sec:loc-gauge-sample} we evaluate the defect free energy for 3d chiral multiplets in 5d gauge theories in the planar limit. Conformal defects and $F$-maximization are discussed in sec.~\ref{sec:CFE} and  defect RG flows in sec.~\ref{sec:M1FE}.
Before proceeding we pause to clarify our conventions for the R-charges.

\bigskip
\textbf{R-charges:} We denote $\mathfrak{u}(1)_R$ the Cartan of the 5d R-symmetry algebra $\mathfrak{su}(2)_R$, and $\mathfrak{u}(1)_r$ is the 3d R-symmetry. We use the conventions in which a supermultiplet in the spin-$q_R$ representation of $\mathfrak{su}(2)_R$ transforms with R-charge $q_R \in \frac{1}{2} \mathbb{Z}$, so in particular a 5d hypermultiplet has $\mathfrak{u}(1)_R$ R-charge 
\begin{equation}
	q_R = \frac{1}{2} ~.
\end{equation}
The $\mathfrak{u}(1)_r$ R-charge of the defect chiral and antichiral multiplets is $q_r \in \RR$. 
The defect R-symmetry arises as the diagonal combination of the Cartan subalgebra $\mathfrak{u}(1)_R$ of the 5d R-symmetry and the $\mathfrak{u}(1)_{\perp}$ rotations transverse to the defect. Additionally, the defect has a vector $\mathfrak{u}(1)_f$ flavor symmetry from the orthogonal linear combination. Denoting the charge of the defect chiral multiplet under the $\mathfrak{u}(1)_{r/f}$ symmetries by $q_{r/f}$,
\begin{align}
	\label{eq:qIRfromqUV}
	q_{r} &= \frac{q_R + q_{\perp}}{2}~,& q_{f} &= \frac{q_R - q_{\perp} }{2} .
\end{align}
One can argue for the normalization by $\frac{1}{2}$ to ensure that $0 \le q_r \le 2$ if $0 \le q_R, q_{\perp} \le 2$. The localization derivation of this relation is a consequence of eq.~\eqref{eq:LrvsLR} below.

By the unitarity bound, the scaling dimension $\Delta$ of a chiral at the conformal point equals the superconformal R-charge. 
The conformal point of the combined 5d/3d system resides in the UV, where $\Delta$ approaches $q_r$ and is not affected by potential mixing with Abelian flavor charges.

\subsection{Localization with 3d defects}\label{sec:localization}

We now give a formal argument for the construction of the 5d/3d matrix models in the next subsection.
The action and supersymmetry transformations of 5d $\cN=1$ theories on the round $S^5$ were constructed in \cite{Hosomichi:2012ek} and used in \cite{Kallen:2012va} to derive a matrix model from localization of the path integral. Here we follow the procedure and conventions of \cite{Kallen:2012va} and include the defect $\cD$.

We begin by adding a 5d hypermultiplet in the fundamental representation of an ambient gauge group $SU(N)$. It consists of the complex scalars $\mathsf{q}_{1,2}$ which form a doublet under the 5d R-symmetry $SU(2)_R$, and the four-complex components spinor $\Psi^{\prime}$. The fields are in the fundamental representation of $Sp(N)$ and decompose into \cite{Hosomichi:2012ek}
\begin{align}
	\mathsf{q}_1 = \frac{1}{\sqrt{2}} \left( \begin{matrix} q \\ \tilde{q} \end{matrix} \right) , \qquad \mathsf{q}_2 = \frac{1}{\sqrt{2}} \left( \begin{matrix} -\tilde{q} ^{\ast} \\ q^{\ast} \end{matrix} \right) , && \Psi^{\prime} = \left( \begin{matrix} \Psi \\ - \mathcal{C} \Psi^{\ast} \end{matrix} \right)
\end{align}
under the diagonal embedding $SU(N) \hookrightarrow Sp(N)$. The field $q$ (resp. $\tilde{q}$) is a complex scalar in the (anti)fundamental representation of $SU(N)$ and $\mathcal{C}$ is the charge conjugation matrix. Additionally, one includes the F-term as in \cite{Hosomichi:2012ek}.

Our defect multiplets always come in chiral/antichiral pairs with equal R-charge, thus their scalar component can be identified with $q, \tilde{q}$ upon restricting their path integral domain to fields supported only on $S^3 \subset S^5$. The spinors $\Psi, - \mathcal{C} \Psi ^{\ast}$, have in total four independent complex components, but are not suited to be associated to 3d chiral multiplets as they are. First, we decompose the 5d charge conjugation matrix $\mathcal{C}$ as 
\begin{equation}
	\mathcal{C} = \left( \begin{matrix} 0 & C \\ -C^{T} & 0\end{matrix}  \right)
\end{equation}
in terms of 2-by-2 matrices, $C$ being the charge conjugation matrix acting on 3d spinors. Then, the Dirac spinor $\Psi$ decomposes into two-complex component spinors as 
\begin{align}
	\Psi &=  \left( \begin{matrix} \psi_1 \\ \psi_2 \end{matrix} \right), & - \mathcal{C} \Psi^{\ast} &=  \left( \begin{matrix} -C \psi_2 ^{\ast} \\ C^{T} \psi_1 ^{\ast} \end{matrix} \right) .
\end{align}
We define $\psi= \psi_1$ and $\tilde{\psi}= -C \psi_2 ^{\ast}$. They have the correct behaviour under gauge and $\mathfrak{u}(1)_R$ R-symmetry transformations to be identified with the superpartners of $q$ and $\tilde{q}$ respectively. We have thus built the desired 3d chiral/antichiral pair out of a 5d hypermultiplet.

The key point is that the localization of \cite{Kallen:2012va} works identically in 3d \cite{Kallen:2011ny} (see also \cite{Lundin:2021zeb} for a uniform treatment of odd-dimensional spheres). In both cases, the localization relies on the Hopf fibration
\begin{equation}\label{eq:Hopffibrations}
\begin{aligned}
	S^1 & \hookrightarrow S^5 \xrightarrow{\quad \pi_{H} ^{({\rm 5d})} \quad } \mathbb{P}^2 \\
	S^1 & \hookrightarrow S^3 \xrightarrow{\quad \pi_{H} ^{({\rm 3d})} \quad } \mathbb{P}^1 .
\end{aligned}
\end{equation}
Let $v_{n}$ denote the Reeb vector field of the Hopf fibration \eqref{eq:Hopffibrations} on $ \mathbb{P}^{n}$, and $\mathcal{L}_{v_n}$ the Lie derivative along $v_n$, for $n=1,2$. In adapted coordinates such that $\theta$ parametrizes the Hopf circle, $v_n = - \partial_{\theta}$. The choice of supercharge consistent with \cite{Kallen:2012va} acting on differential forms, yields \cite{Santilli:2020uht}
\begin{equation}\label{eq:deltasquareloc}
\begin{aligned}
 {\rm 5d} \ : && \delta^2 &= i \mathcal{L}_{v_2} + i \sigma + \Delta^{\scriptscriptstyle (R)} ~, \\
 {\rm 3d} \ : && \delta^2 &= i \mathcal{L}_{v_1} + i \sigma + \Delta^{\scriptscriptstyle(r)} ~, \\
\end{aligned}
\end{equation}
where $\Delta$ is the scaling dimension of the field, and we use the superscript to emphasize that the unitarity bound relates $\Delta$ to the R-charges for $\mathfrak{u}(1)_R$ in 5d and $\mathfrak{u}(1)_r$ in 3d.

\bigskip
We select an $S^3 \subset S^5$ by requiring that the embedding of its Reeb vector field $v_1$ equals the Reeb vector field $v_2$ of the ambient $S^5$. Then, setting $\pi_H = 1- \iota_{v_2}$, we specify a map $\varpi$ such that it defines a commutative diagram from \eqref{eq:Hopffibrations}:
\begin{equation*}
\begin{tikzpicture}
\node (s5) at (-2,1) {$S^5$};
\node (s3) at (-2,-1) {$S^3$}; 
\node (p2) at (2,1) {$\mathbb{P}^2$};
\node (p1) at (2,-1) {$\mathbb{P}^1$}; 

\path[->] (s5) edge node[anchor=south] {$\pi_{H} $} (p2);
\path[->] (s5) edge node[anchor=east] {$\varpi $} (s3);
\path[->] (p2) edge node[anchor=west] {$\varpi \circ \pi_H $} (p1);
\path[->] (s3) edge node[anchor=north] {$\pi_H \circ \varpi $} (p1);
\end{tikzpicture}
\end{equation*}
In more physical terms, the choice of Hopf fibre in $S^5$ fixes the supercharge used for localization. We then identify the 3d $\mathcal{N}=2$ supercharges used to localize on $S^3$ with ambient ones, equivalent to specifying an embedding of (part of) the 3d $\mathcal{N}=2$ superalgebra into the 5d $\mathcal{N}=1$ superalgebra.\par
To set the notation, let $\Nb$ be the normal bundle to $\mathbb{P}^1$ inside $\mathbb{P}^2$, so that we have the short exact sequence 
\begin{equation}
\label{eq:SESeqN}
	0 \to T \mathbb{P}^1 \to \left. T \mathbb{P}^2 \right\rvert_{\mathbb{P}^1}\to \Nb \to 0 ~.
\end{equation}
Let also $\overline{\mathscr{K}}_{\mathbb{P}^{n}}$ be the anticanonical bundle of $\mathbb{P}^{n}$, $n=1,2$. Besides, $\mathscr{L}^{(R)}$ and $\mathscr{L}^{(r)}$ denote the R-symmetry bundles in 5d and 3d respectively. The pushforward and pullback of a map $f$ will be $f_{\ast}$ and $f^{\ast}$ , as usual.\par
Our strategy is to (i) twist the ambient fields using $\mathscr{L}^{(R)}$, (ii) twist the defect fields using $\mathscr{L}^{(r)}$, and (iii) make these two steps compatible. For this, we need a suitable choice of background connections $A^{(R)}$ and $A^{(r)}$ for the R-symmetry bundles. Taking inspiration from e.g. \cite{Closset:2016arn,Closset:2017zgf}, we pick (locally)
\begin{align}\label{eq:dAR}
d A^{(R)} = - \frac{1}{2} c_1 (\mathbb{P}^2 ) ~, && d A^{(r)} = - \frac{1}{2} c_1 (\mathbb{P}^1 ) ~,
\end{align}
so to cancel the spin connections in all covariant derivatives. In terms of line bundles, the curvatures \eqref{eq:dAR} translate into the choice
\begin{align}\label{eq:LineKbundle}
\mathscr{L}^{(R)} = \sqrt{\pi_H^{\ast} \overline{\mathscr{K}}_{\mathbb{P}^2}} ~, && \mathscr{L}^{(r)} = \sqrt{ (\pi_H \circ \varpi)^{\ast} \overline{\mathscr{K}}_{\mathbb{P}^1}} ~.
\end{align}
We thus have 
\begin{equation}\label{eq:c1r3d}
	c_1 ((\pi_{H}\circ \varpi )_{\ast} \mathscr{L}^{(r)}) = - \frac{1}{2} c_1 (\mathbb{P}^1)
\end{equation} 
and 
\begin{align}
	c_1 (\pi_{H, \ast} \mathscr{L}^{(R)}) & = - \frac{1}{2} \left( \dim_{\mathbb{C}} T^{(1,0)} \mathbb{P}^2 -1 \right) c_1 (\mathbb{P}^2) \nonumber \\
	&= - \frac{1}{2} \left( 3 -1 \right) \left[ c_1 (\mathbb{P}^1) + c_1(\mathscr{N})  \right] = \  - c_1 (\mathbb{P}^1) - c_1(\mathscr{N}) ~, \label{eq:c1Pc1R}
\end{align}
where the first equality follows from \eqref{eq:LineKbundle} using the Chern class of the anticanonical bundle, and in the second equality we have used \eqref{eq:SESeqN}. Comparing \eqref{eq:c1r3d} with \eqref{eq:c1Pc1R}, the compatibility condition on the R-symmetry bundles is:
\begin{equation}
	2 c_1 \left( (\varpi \circ \pi_{H}))^{\ast} ((\pi_{H}\circ \varpi)_{\ast}  \mathscr{L}^{(r)}) \right) = c_1 (\pi_{H,\ast} \mathscr{L}^{(R)}) + c_1 (\mathscr{N}) 
\end{equation}
expressed in terms of line bundles on $\mathbb{P}^2$. Omitting the various projections, the latter equation is written less precisely but more cleanly as 
\begin{equation}\label{eq:LrvsLR}
	2 c_1 \left(\mathscr{L}^{(r)} \right) = c_1 \left(\mathscr{L}^{(R)} \right) + c_1 \left(\mathscr{N}\right) ~,
\end{equation}
establishing a quadratic relation between $\mathscr{L}^{(r)}$ and $ \mathscr{L}^{(R)} \otimes \mathscr{N}$. The factor of 2 on the left-hand side of \eqref{eq:LrvsLR} means that we should replace the R-charge $2q_r= q_R + q_{\perp}$ in the localization formula. It hence leads to eq.~\eqref{eq:qIRfromqUV} for the R-charges, and will play an important role.\par
We can now apply the localization machinery. The supersymmetry operator is the five-dimensional one of \cite{Kallen:2012va}, whose square is in \eqref{eq:deltasquareloc}. It acts equally on ambient and defect fields, the latter being differential forms of lower degree. When acting on forms of definite degree, we write \eqref{eq:deltasquareloc} schematically as 
\begin{align}
	\delta^2 \alpha = i \mathsf{L}_{\Omega ^{\bullet}} (\sigma) \cdot \alpha ~, && \alpha \in \Omega ^{\bullet} (S^3, \mathfrak{g}) ~.
\end{align}
Then, after gauge fixing, the calculation of the one-loop determinants on the defect reduces to evaluate \cite{Santilli:2020uht}
\begin{equation}\label{eq:ratio1loopdets}
	\cZ_{\cD} ^{\text{1-loop}} = \sqrt{ \frac{ \det  i \mathsf{L}_{\Omega_{\mathrm{h}} ^{(0,1)}} (\sigma) }{ \left(  \det  i \mathsf{L}_{\Omega_{\mathrm{h}} ^{0}} (\sigma) \right)^2 } }  \sqrt{ \frac{ \det  i \mathsf{L}_{\Omega_{\mathrm{h}} ^{(0,1)}} (-\sigma) }{ \left(  \det  i \mathsf{L}_{\Omega_{\mathrm{h}} ^{0}} (- \sigma) \right)^2 } } ~.
\end{equation}
The two factors come from the chiral and antichiral multiplet, respectively, and the subscript in $\Omega^{\bullet}_{\mathrm{h}}$ indicates horizontal forms. That is, $\alpha \in \Omega^{\bullet}_{\mathrm{h}}$ if $\alpha \in \Omega^{\bullet}$ such that $\iota_v \alpha =0$, equivalently, $\pi_H \alpha = \alpha$. The derivation and evaluation of \eqref{eq:ratio1loopdets} via an index theorem is given in \cite{Santilli:2020uht}.\par
The final answer for the one-loop determinant $\cZ_{\cD} ^{\text{1-loop}}$ agrees with the one-loop determinant of a genuinely 3d chiral/antichiral pair found in \cite{Jafferis:2010un,Hama:2010av}. This proves that the contribution of $\cD$ to the $S^5$ partition function equals the inclusion of a 3d chiral/antichiral multiplet pair in the fundamental representation of $SU(N)$ in the matrix model.\par

\bigskip
\textbf{Remarks:} We conclude this part with a few remarks. First, we considered specific defects comprising 3d chiral multiplets minimally coupled to a 5d vector multiplet, possibly with a superpotential. The localization problem for arbitrary defects is outside the scope of this work. Notwithstanding, the derivation extends to more general 5d gauge theories, also beyond the linear quivers of interest here. As a simple application, codimension-2 defects in 5d maximally supersymmetric Yang--Mills theory are studied in app.~\ref{app:MSYM}.

The codimension-2 defect contribution is evaluated in the background of the trivial connection on $S^5$, i.e. discarding 5d instantons. This is sufficient for our purposes, since instantons decouple in the planar limit we consider below.

As happens already for the localization of a 3d chiral multiplet coupled to a gauged 3d vector multiplet, the one-loop determinant has poles in the complex plane. The localization procedure holds for $0< \Delta < 2$, with the one-loop determinant singular at $\Delta \in 2 \mathbb{Z}$, where the poles pinch the integration contour. These singularities are ubiquitous in supersymmetric localization and are due to the appearance of chiral multiplet zero-modes \cite{Hori:2013ika,Closset:2017zgf}. 

Finally, the machinery deployed here can be adapted to codimension-2 defects engineered by a 5d $\cN=1$ hypermultiplet coupled to 7d maximally supersymmetric Yang--Mills, whose perturbative partition function on $S^7$ was analyzed in \cite{Minahan:2015jta}.

\subsection{Matrix models}\label{sec:localization-F}

We now discuss the matrix models for the defect free energy of codimension-2 defects in 5d linear quiver gauge theories.
The generic form of the quivers is shown in fig.~\ref{fig:quiver}, to fix notation. Our explicit examples will be gauge theory deformations of the $T_N$, $+_{N,M}$ and $Y_N$ theories in fig.~\ref{fig:pqwebs}.
The 3d defects we consider here are defined in these gauge theories. The infinite coupling limit, leading to the 5d SCFTs, is understood to be taken at the end. However, the general derivations in this subsection are independent of the gauge coupling.

The matrix models describing 5d quiver gauge theories were derived in \cite{Kallen:2012cs,Kallen:2012va,Lockhart:2012vp,Kim:2012qf,Closset:2022vjj}. Here we leave the discussion of the ambient CFT schematic and focus on the defect contributions; all we need from the ambient CFT are the saddle points dominating the matrix models in the planar limit. 
The $S^5$ partition function of a 5d quiver gauge theory as in fig.~\ref{fig:quiver} can be written as
\begin{align}\label{eqn:partfuncnoD}
\cZ_{S^5}  = \left[ \prod_{t=1}^{L} \prod_{j=1}^{N_t} \int_{-\infty}^{\infty} d \lambda_j^{(t)} \right] e^{-S(\lambda)}~,
\end{align}
where $\lambda_j^{(t)}$ is the $j^{\rm th}$ eigenvalue associated with the $t^{\rm th}$ gauge node and $S(\lambda)$ denotes the integrand, including one-loop determinants and classical prepotential of the 5d fields. The constraint $\prod_{t=1}^L\delta \left( \sum_{j=1}^{N_t} \lambda_j ^{(t)}\right)$ is implicit in \eqref{eqn:partfuncnoD}. We denote by $F_{S^5} \equiv F_{S^5} [\emptyset ]$ the free energy of the 5d gauge theory without defect.

To compute the contribution of the defect $\cD$ to the sphere partition function, we need to incorporate the one-loop determinant of the 3d chiral multiplets into the matrix model for the 5d theory. Following the discussion in sec.~\ref{sec:localization}, we can simply import the defect one-loop determinant from the purely 3d results \cite{Jafferis:2010un,Hama:2010av}.\footnote{The same logic was used in \cite{Robinson:2017sup,Komatsu:2020sup,Beccaria:2022bjo} for 3d defects in 4d $\mathcal{N}=4$ super-Yang--Mills.} 
We arrive at
\begin{align}\label{eqn:generalpartfunc}
	\cZ_{S^5} \left[ \cD \right] = \left[ \prod_{t=1}^{L} \prod_{j=1}^{N_t} \int_{-\infty}^{\infty} d \lambda_j^{(t)} \right]e^{-S(\lambda)}\cZ_{\cD} ^{\text{1-loop}}[t_0]~,
\end{align}
where $t_0$ indicates the gauge group under which the defect is charged.
With $(\Delta, \tilde{\Delta})$ denoting the scaling dimensions of the chiral/antichiral pair, the contribution of $\cD$ to the matrix model is
\begin{align}
\cZ_{\cD} ^{\text{1-loop}}[t_0] &=  \prod_{j=1}^{N_{t_0}} e^{-F^{(3d)}_H\left(\lambda_j^{(t_0)}\right)},& \ F^{(3d)}_H(\lambda)&=-\ell(1-\Delta+i\lambda)-\ell(1-\tilde{\Delta}-i\lambda)~.
\end{align}
The function $\ell (z)$, defined in \cite{Jafferis:2010un}, satisfies
\begin{align}
	\ell(z)&=- z\ln \left(1-e^{2\pi i z}\right)+\frac{i}{2}\left(\pi z^2+\frac{1}{\pi}\Li_2\left(e^{2\pi i z}\right)\right)-\frac{i\pi}{12}~, \\
	\partial_z \ell (z)&=-\pi z \cot(\pi z) , \qquad \quad \qquad \ell \left( \frac{1}{2} +i\lambda \right) + \ell \left( \frac{1}{2} - i\lambda \right) = - \ln(2\cosh(\pi\lambda))~. \nonumber
\end{align}
Denoting by
\begin{align}
	F_{S^5} \left[ \cD \right] = - \ln \cZ_{S^5} \left[ \cD \right]  
\end{align}
the sphere free energy of the 5d theory coupled to the defect, the defect free energy captures the contribution of $\cD$ and is defined as
\begin{equation}
	F_{\cD} = F_{S^5} \left[ \cD \right]  - F_{S^5} \left[ \emptyset \right] .
\end{equation}

The planar limit of the 5d quiver in fig.~\ref{fig:quiver} amounts to taking the gauge group ranks $\left\{ N_t \right\}$ as well as the length $L$ of the quiver large. This long quiver limit was studied in \cite{Fluder:2018chf,Uhlemann:2019ypp}.
The contribution due to $\cD$ is subleading compared to the 5d integrand in \eqref{eqn:generalpartfunc}, and hence does not affect the saddle point equations. Therefore, the defect free energy can be obtained in the planar limit by evaluating the extra insertion on the saddle point eigenvalue distribution derived in \cite{Uhlemann:2019ypp}. Explicitly:
\begin{align}\label{eq:Fdefectsaddle}
 F_{\cD} &=-\ln \left[\prod_{j=1}^{N_{t_0}} e^{-F^{(3d)}_H\left(\lambda_j^{(t_0)}\right)}\right] \Bigg\vert_{\rm saddle}  =\sum_{j=1}^{N_{t_0}} F^{(3d)}_H\left(\lambda_j^{(t_0)}\right)\Big\vert_{\rm saddle} ~.
\end{align}
Using the eigenvalue densities $\rho_{t_0}$ for the $t_0^{\rm th}$ gauge node, 
\begin{align}
	 \rho_{t_0}(\lambda) =\frac{1}{N_{t_0}} \sum_{j=1}^{N_{t_0}} \delta \left(\lambda - \lambda_j^{(t_0)}\right)~,
\end{align}
expression \eqref{eq:Fdefectsaddle} becomes
\begin{align}\label{eq:delta-F-0}
 F_{\cD} &=N_{t_0} \int d\lambda\, \rho_{t_0}(\lambda) F^{(3d)}_H\left(\lambda\right)~.
\end{align}
For the long quiver theories, the eigenvalues are captured by a function of two variables, 
\begin{align}
 \rho(z,\lambda)\equiv \rho_{\lfloor zL \rfloor }(\lambda)~.
\end{align}
Here $\lfloor zL \rfloor$ means the integer part of $zL$ and indicates the gauge node. We will also write $N(z) \equiv N_{\lfloor zL \rfloor }$. The scaling of the eigenvalues determined in \cite{Uhlemann:2019ypp} for the round $S^5$ is 
\begin{align}\label{eq:lambda-scaling}
 \lambda&=3Lx~,
 &
 \hat\rho(z,x)&=3L\rho(z,3L x)~,
\end{align}
with $x$ of $\mathcal O(1)$ and $\hat\rho$ a properly normalized distribution for the $x$. Denoting $z_0 = \frac{t_0}{L}$, this leads to 
\begin{align}\label{eq:delta-F}
F_{\cD} (z_0)&=N(z_0) \int dx\, \hat\rho(z_0,x)F^{(3d)}_H(3 L x)~.
\end{align}
Furthermore, one can use that for large argument
\begin{equation}\label{eq:F3dDDbar}
F^{(3d)}_H(\lambda) \approx \pi(2-\Delta - \tilde{\Delta} )\lvert\lambda\rvert ~.
\end{equation}
If the flavor symmetry on the defect is preserved, we have $\Delta = \tilde{\Delta}$. Then \eqref{eq:delta-F} simplifies to
\begin{align}\label{eq:delta-F-2}
 F_{\cD}(z_0) &=6\pi L(1-\Delta) N(z_0) \int dx\, \hat\rho(z_0,x)|x|~.
\end{align}

\subsection{Gauge theory case studies}\label{sec:loc-gauge-sample}

We now discuss 3d flavor defects in a sample of gauge theories. From the gauge theory perspective we can add 3d flavors to any gauge node in a long quiver gauge theory. This yields a one-parameter family of defects for each 5d gauge theory.

\begin{figure}
	\centering
	\subfigure[][]{\label{fig:FDHBflow-1}
		\includegraphics[width=0.3\textwidth]{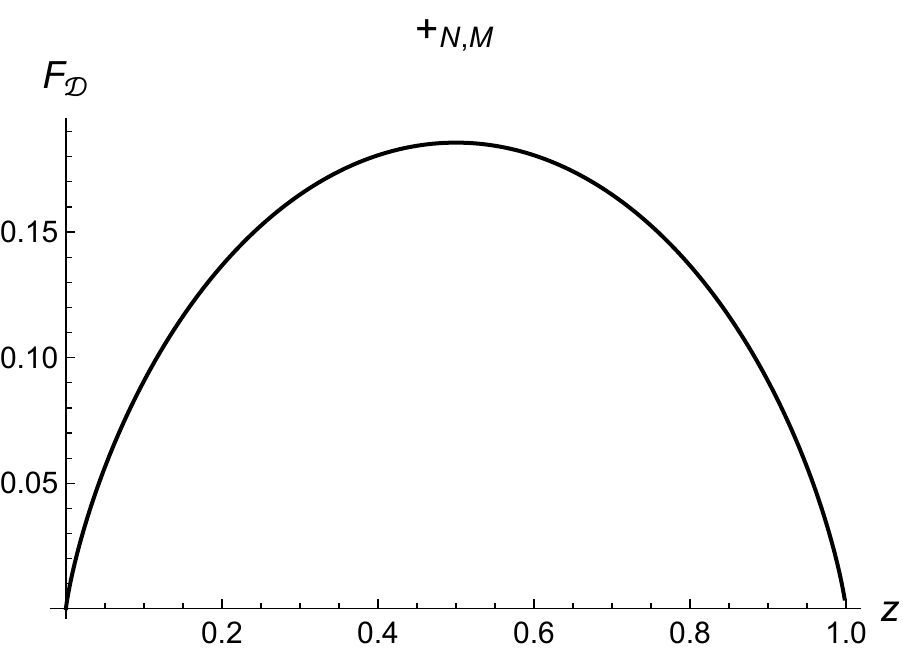}
	}\hskip 2mm
	\subfigure[][]{\label{fig:FDHBflow-2}
		\includegraphics[width=0.3\textwidth]{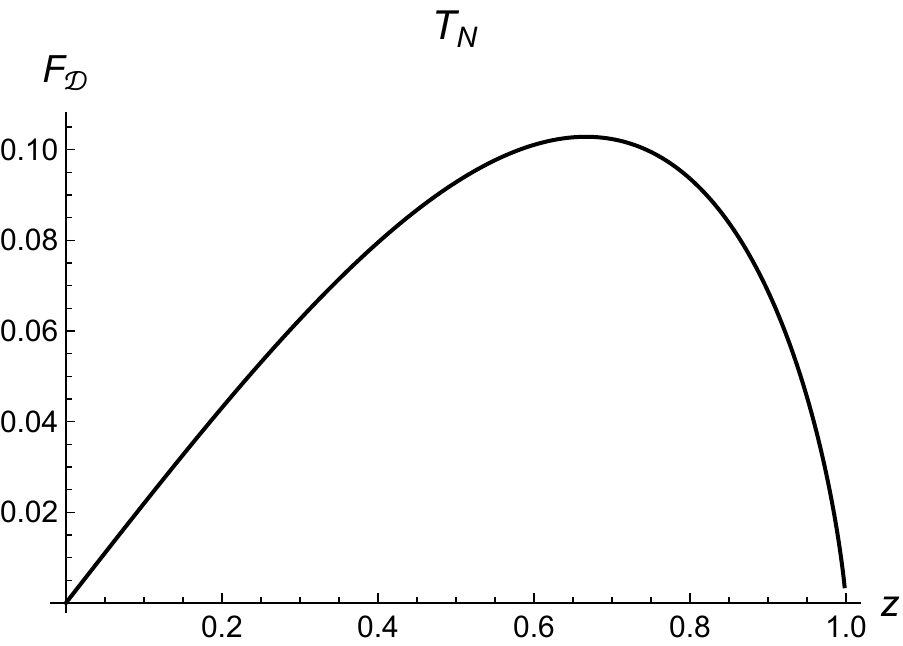}
	}\hskip 2mm
	\subfigure[][]{\label{fig:FDHBflow-3}
		\includegraphics[width=0.3\textwidth]{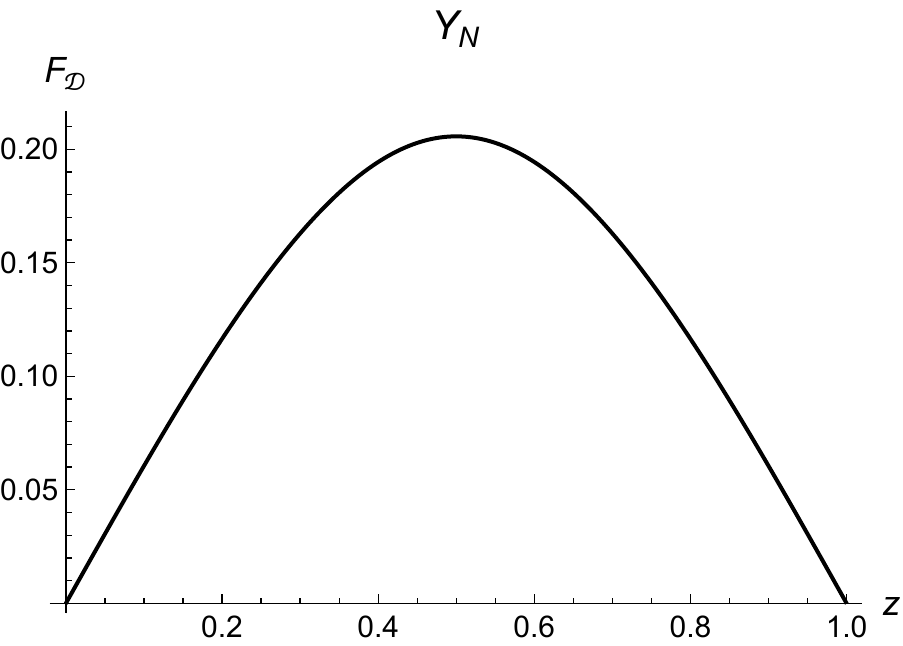}
	}
	\caption{$F_{\cD}/[6 \pi (1-\Delta)NL]$ as a function of $z$. (a) $+_{N,M}$, (b) $T_N$, (c) $Y_N$.}
	\label{fig:FDHBflow}
\end{figure}

\textbf{\boldmath{$+_{N,M}$} theory:} We start with the $+_{N,M}$ theory. The gauge theory deformation has $L=M-1$ gauge nodes and is described by the quiver
\begin{align}
	[N]-\underbrace{SU(N)-\cdots -SU(N)}_{SU(N)^{M-1}}-[N]~.
	\label{eq:D5NS5-quiver}
\end{align}
All gauge nodes are balanced and have Chern--Simons level zero.
The saddle point eigenvalue distributions $\hat\rho(z,x)$ for the $+_{N,M}$ theory with the gauge couplings taken to infinity is \cite{Uhlemann:2019ypp}
\begin{align}
	\hat\rho_s(z,x)&=\frac{4\sin (\pi  z) \cosh \left(2 \pi x\right) }{\cosh \left(4 \pi x\right)-\cos (2 \pi  z)}~.
\end{align}
From (\ref{eq:delta-F-2}) we find the defect free energy for a 3d defect associated with the gauge node $z_0=t_0/L$ as
\begin{align}
	F_{\cD}(z_0)&=\frac{6MN}{\pi}(1-\Delta )\Im\left[\Li_2\left(e^{i \pi  z_0}\right)-\Li_2\left(-e^{i \pi 
		z_0}\right)\right].
\end{align}
This function is plotted in fig.~\ref{fig:FDHBflow-1}. The ends of the quiver correspond to $z=0,1$, and the plot clearly shows the symmetry under $z\rightarrow 1-z$.

\textbf{\boldmath{$T_N$} theory:} We next turn to the $T_N$ theories. The gauge theory deformation is given by 
\begin{align}
	\label{TNquiver}
	[2]-SU(2)-SU(3)-\cdots-SU(N-1)-[N] ~.
\end{align}
The Chern--Simons levels are again all zero and the nodes are balanced.
The saddle point eigenvalue distribution is
\begin{align}\label{eq:TN-saddle}
	\hat\rho_s(z,x)&=\frac{\sin (\pi  z)}{z}\frac{1}{\cosh \left(2\pi x\right)+\cos (\pi  z)}~.
\end{align}
The defect free energy for a defect at the node $z_0=t_0/L$ is
\begin{align}
	F_{\cD}(z_0)&=\frac{6 N^2}{\pi} (1-\Delta ) \Im\Li_2\left(-e^{-i \pi  z_0}\right)~.
\end{align}
A plot is shown in fig.~\ref{fig:FDHBflow-2}. Again, the defect contribution vanishes when the 3d defect is associated with a node near one of the ends of the quiver but is positive in between.

\textbf{\boldmath{$Y_N$} theory:} The $Y_N$ theory can be realized by taking two copies of $T_N$ and identifying and gauging the flavor node. It admits the gauge theory representation
\begin{align}\label{eq:YN-quiver}
	SU&(2)-SU(3)-\cdots-SU(N\,{-}\,1)-SU(N)^{}_{\pm 1}-SU(N\,{-}\,1)-\cdots-SU(3)-SU(2)
	\nonumber\\
	&\ \vert \hskip 130mm \vert
	\\  \nonumber
	&[2] \hskip 127mm [2]
\end{align}
The central node has Chern--Simons level $\pm 1$, indicated by the subscript. We set it to $+1$.
There is an S-dual gauge theory description given by the balanced quiver 
\begin{align}\label{eq:YN-quiver-S}
	SU(2)-SU(4)-SU(6)-\cdots-SU(2N-2)-[2N] ~.
\end{align}
This quiver differs from that of the $T_N$ theory in \eqref{TNquiver} by a rescaling of the gauge node ranks and the $\mathcal O(1)$ flavors on the left end. As a result, the saddle point eigenvalue distribution for the S-dual quiver is the same as \eqref{eq:TN-saddle} \cite{Uhlemann:2019ypp}, and $F_{\cD}$ computed for \eqref{eq:YN-quiver-S} is simply twice the value for $T_N$.
The natural 3d flavor defects from the gauge theory perspective differ: the quiver in (\ref{eq:YN-quiver-S}) has $N-1$ nodes to which 3d chiral multiplets can be coupled, while the quiver in (\ref{eq:YN-quiver}) has $2N-1$ nodes. However, since there is only one conformal D3-brane embedding in the supergravity dual of the $Y_N$ SCFT, we expect conformal defects to agree.

We focus on the description in \eqref{eq:YN-quiver} in the following, which, unlike the  $+_{N,M}$ and $T_N$ quivers, has an unbalanced node and a Chern-Simons term. The Chern-Simons term in particular breaks symmetry under $\lambda\rightarrow -\lambda$.
The saddle point distribution is given by
\begin{align}
	\hat\rho_s(z,x)&=\frac{1}{1-|1-2z|}\cdot\frac{2}{\sqrt{-1-4e^{-4\pi x+2\pi iz}}}+\rm{c.c.}
\end{align}
and has bounded support at the central node, where it is non-zero for $2\pi x\leq \ln 2$.
A plot of the defect contribution to the free energy is shown in fig.~\ref{fig:FDHBflow-3}. Integrating \eqref{eq:delta-F-2} with this eigenvalue density, we arrive at 
\begin{align}
	F_{\cD}(z_0)=\frac{12N^2}{\pi} (1-\Delta ) \Im \left[ \Li_2\left(\frac{4e^{i2\pi z_0}}{(1+\sqrt{1+4e^{i2\pi z_0}})^2}\right) + \ln \left(\frac{1+\sqrt{1+4e^{i2\pi z_0}}}{2}\right) \right]~.
\end{align}
To simplify this expression, factors of $\sqrt{-1}$ have been treated assuming $0\le z_0 \le\frac{1}{2}$. The expression for $ \frac{1}{2}\le z_0\le 1$ differs by an overall minus sign.

\subsection{Defect F-maximization}\label{sec:CFE}

With the defect free energies for 3d defects attached to arbitrary gauge nodes in the $+_{N,M}$, $T_N$ and $Y_N$ theories in hand, the natural next question is whether these results capture the conformal defect free energies computed in sec.~\ref{sec:holography}. We find that the conformal defect free energies can be recovered by an extremization of the defect free energy over the gauge node which the defect is attached with:
\begin{align}
	F_{\cD}^{\text{conf.}}&=\max_{z_0}F_{\cD}(z_0)~.
\end{align}
This leads to the following values for our sample theories:
\begin{align}\label{eq:z0-plus-TN}
	+_{N,M}:& \qquad
	z_0=\frac{1}{2}
	&
	T_N:& \qquad z_0=\frac{2}{3}
	&
	Y_N:&\qquad z_0=\frac{1}{2} ~. 
\end{align}
The value for the $Y_N$ theory is for the quiver in (\ref{eq:YN-quiver}); the S-dual quiver in (\ref{eq:YN-quiver-S}) would lead to $z_0=2/3$.
We note that the values of $z_0$ match the gauge nodes identified independently in eq.~(\ref{eq:gauge-nodes-sugra}).
With these values we find the free energy for conformal defects as
\begin{align}
	+_{N,M}:&&  F_{\mathcal{D}} &=\frac{12C}{\pi}(1-\Delta) MN~, && \notag \\
	T_N:&& F_{\mathcal{D}} &= \frac{6}{\pi}(1-\Delta)\Im {\rm Li}_2\big(e^{i\pi/3}\big)N^2~, \notag &&
	\\
	Y_N:&& F_{\mathcal{D}} &= \frac{12}{\pi}(1-\Delta)\Im {\rm Li}_2\big(e^{i\pi/3}\big)N^2~. && \label{eq:3FDCFT}
\end{align}
Note that $F_{\mathcal{D}} ^{Y_N} = 2F_{\mathcal{D}} ^{T_N}$. This is in line with the discussion below (\ref{eq:YN-quiver-S}): both quivers for the $Y_N$ theory should lead to the same defect free energy for the conformal defect.

The results for $F_{\cD}$ still depend on the R-charge $\Delta$ through the factor $(1- \Delta)$, as a consequence of the expression \eqref{eq:delta-F-2}. Recall that $\Delta$ equals the superconformal R-charge, which typically mixes $\mathfrak{u}(1)_r$ with Abelian global symmetries upon flowing to the IR. However, the fixed point of interest to us for the combined 5d/3d system is in the UV. We therefore set $\Delta=q_r$ to its UV value, which from \eqref{eq:qIRfromqUV} is $\Delta = \frac{1}{4}$. All results \eqref{eq:3FDCFT} then match the holographic results of sec.~\ref{sec:conf-hol} for $\Delta=\frac{1}{4}$.

When comparing to the supergravity results, the value of $z_0$ for the $+_{N,M}$ and $Y_N$ theories in \eqref{eq:z0-plus-TN} can be understood based on symmetry: The $+_{N,M}$ and $Y_N$ 5-brane junctions possess a $\mathbb{Z}_2$ symmetry acting as horizontal reflection. This $\mathbb{Z}_2$ symmetry is also manifest in the gauge theories, acting as reflection across the central node, $z \mapsto 1-z $. The value $z_0= \frac{1}{2}$ is the fixed point of this symmetry, and adding a 3d defect in the form of a 3d chiral/antichiral pair at the central gauge node preserves this $\ZZ_2$ symmetry. This is in line with the conformal defect D3-brane being at a location which is invariant under the $\ZZ_2$ symmetry in the supergravity solutions.

\bigskip
\textbf{Defect Higgs branch flows:}
The above notion of $F$-maximization can be motivated by the $F$-theorem for 3d defects as follows.
In the brane engineering of the  3d defects there is a deformation displacing the D3-brane from the conformal point along a D5-brane (fig.~\ref{fig:massive-D3-plus}). We described this in sec.~\ref{sec:defects} as a form of defect Higgs branch deformation.
Starting with a defect at the gauge node $t_0$ as in \eqref{eq:z0-plus-TN}, other values of $t_0$ can be reached via successive defect Higgs branch flows.
The plots in fig.~\ref{fig:FDHBflow} show that along such flows the defect free energy $F_{\cD}$ decreases if the conformal seed defect is at the maximum of the defect free energy.

For defect RG flows which preserve an automorphisms of the quiver, e.g.\ the $\ZZ_2$ reflection of the $+_{N,M}$ quiver in (\ref{eq:D5NS5-quiver}),
the analysis applies directly to the orbifold theories constructed in \cite{Bergman:2015dpa,Uhlemann:2019lge}. With two copies of $\cD$ placed at $z_0=\frac{1}{2}$, Higgs branch flows that preserve the $\mathbb{Z}_2$ automorphism of the quiver lead to defects charged under the $\mathbb{Z}_2$-symmetric gauge nodes at $z_0 \pm \varepsilon$, $0< \varepsilon < \frac{1}{2}$. The combined system descends to a defect of the orbifold. 
For the S-folds of \cite{Acharya:2021jsp,Kim:2021fxx,Apruzzi:2022nax} this is less straightforward, since the automorphisms are typically not visible in gauge theory deformations.

\subsection{Defect RG flows}
\label{sec:M1FE}

We now turn to defect mass deformations. This will allow us to show that the match between the defect free energies obtained from the field theory $F$-maximization in sec.~\ref{sec:CFE} on the one hand, and the supergravity predictions of sec.~\ref{sec:holography} on the other, can be extended from conformal defects to a match between certain deformations and the resulting RG flows. We will also discuss the phase transition found for the $Y_N$ theory in sec.~\ref{sec:AdScase} from the field theory perspective.

We consider a relevant deformation of the defect by a uniform real mass term $m$ for the 3d chiral multiplets. It is associated to the vector $\mathfrak{u}(1)$ flavor symmetry that rotates the chiral and antichiral multiplets on the defect. 
The explicit form of the mass term with the associated (imaginary) curvature coupling can be found e.g.\ in \cite[(2.12)]{Willett:2016adv}.
The mass term preserves both $\mathfrak{u}(1)$ symmetries preserved by the D3-brane RG flows discussed in sec.~\ref{sec:BPS-flow}.
The effect in the matrix model is to shift the coupling to the vector multiplet of $SU(N_{t_0})$ according to 
\begin{equation}
	\lambda_j \longrightarrow \lambda_j + m~, \qquad \qquad j=1, \dots, N_{t_0}~.
\end{equation}
Eq.~\eqref{eq:delta-F-0} becomes 
\begin{align}
	F_{\cD} &=N_{t_0} \int d\lambda\, \rho_{t_0}(\lambda) F_H ^{(3d)} (\lambda+m)~.
\end{align}
After the same replacements which led to \eqref{eq:delta-F-2}, we find
\begin{align}\label{eqn:FDmassive}
	F_{\cD} &=6\pi (1-\Delta) L N_{t_0} \int dx\, \hat\rho(z_0,x)|x+\mu|~,
	&
	m=3L\mu~.
\end{align}
The task is to evaluate \eqref{eqn:FDmassive}, with $\hat{\rho}$ and $z_0$ as in the conformal case. To perform these integrals systematically, it is convenient to take two derivatives with respect to $\mu$, to produce a $\delta$ function in the integrand. Then, the result is integrated twice with respect to $\mu$. The integration constants can be fixed using the massless free energy and the derivative with respect to $\mu$ at $\mu=0$.

\bigskip

We focus on defects associated with the gauge nodes singled out by the defect $F$-maximization of sec.~\ref{sec:CFE}, which led to a match between gauge theory and supergravity results for the conformal defects.
For the $+_{N,M}$ and $T_N$ theories we find
\begin{align}
	+_{N,M}:&&  F_{\cD}&=
	6 \pi (1-\Delta) M N \left( \mu +\frac{2}{\pi^2} \Im\Li_2\left(e^{-2 \pi \mu +i\pi/2 }\right)\right)~, &&
	\label{massiveFDpmn}\\
	T_N:&&
	F_{\cD}&=
	4 \pi (1-\Delta) N^2 \left( \mu + \frac{3}{2\pi^2}\Im\Li_2\left(e^{- 2 \pi \mu+i\pi/3 }\right)\right)~.  \label{massiveFDtn}
\end{align}
It follows from the properties of the polylogarithm that the defect free energies are symmetric under $\mu \mapsto - \mu$, and at $\mu=0$ the conformal results of sec.~\ref{sec:CFE} are recovered.

For the $+_{N,M}$ theory the conformal defect is associated with the central gauge node, which could be motivated independently by the $\ZZ_2$ symmetry under reflection of the quiver. The 3d mass deformation preserves this symmetry. One may therefore expect a link to deformations corresponding to vertical displacements of the D3-branes in the string theory construction. Indeed, comparing the field theory result in (\ref{massiveFDpmn}) to the supergravity result for the defect free energy in (\ref{eq:SD3-plus}),  we find a match if the 3d mass $\mu$ in (\ref{massiveFDpmn}) is identified with the  deformation parameter $\mu$ in (\ref{eq:SD3-plus}), which is real for the $\ZZ_2$-symmetric embeddings.
Real $\mu$ was precisely the condition for the D3-brane to preserve the $\ZZ_2$ symmetry. There is no phase transition: from the string theory perspective the D3 can slide along the NS5 branes without constraint, and from the field theory perspective the saddle point eigenvalue density has unbounded support at the central node.

The picture for the $T_N$ theory is similar. There is no symmetry argument to identify the gauge node, but the supergravity discussion in sec.~\ref{sec:gauge-theory-connecttion} and the field theory $F$-maximization of sec.~\ref{sec:CFE} identify the same gauge node, leading to a match of the conformal defect free energies. The field theory result for the mass-deformed defect in (\ref{massiveFDtn}) likewise matches the (unambiguous dilogarithm terms of the) supergravity result in (\ref{eq:SD3-TN-IR}) if $\mu$ is identified. There is again no phase transition.

\bigskip
The discussion for the $Y_N$ theory is more interesting. The supergravity discussion in $Y_N$ identified a phase transition in the defect free energy. Symmetry considerations, the discussion of sec.~\ref{sec:gauge-theory-connecttion} and the field theory $F$-maximization of sec.~\ref{sec:CFE} all link the conformal D3-brane defect to the central gauge node. Indeed, at this node the saddle point eigenvalue density $\hat\rho$ has bounded support, so we can expect a richer picture.
For the second derivative of $F_{\mathcal{D}}$ we find
\begin{align}\label{eq:dsquareFDYN}
	\partial^2_{\mu} F_{\mathcal{D}}^{Y_N} = \begin{cases}0 & \mu< - \frac{\ln 2}{2\pi} ~, \\ \frac{48 \pi}{\sqrt{4e^{4\pi \mu } -1}} ~(1-\Delta) N^2  & \mu> - \frac{\ln 2}{2\pi}~.\end{cases} 
\end{align}
Scheme dependence affects terms linear in $\mu$, hence this expression is free from ambiguities. Furthermore, \eqref{eq:dsquareFDYN} shows that the mass-deformed $Y_N$ theory, in which the eigenvalue density has bounded support in one direction, undergoes a phase transition when the mass crosses the maximal eigenvalue. The phase transition is second order, and takes place at the critical value of the mass $\mu_c = - \frac{1}{2\pi} \ln 2 $. The critical behaviour is controlled by the jump of the second derivative at the critical locus. Near the transition point, $\partial^2_{\mu} F_{\mathcal{D}}^{Y_N} $ vanishes if $\mu < \mu_c$, while for $\mu>\mu_c$
\begin{align}
	\partial^2_{\mu} F_{\mathcal{D}}^{Y_N} \big\vert_{\mu>\mu_c}=  24 \sqrt{\pi} (1-\Delta) N^2 ~ \left( \mu-\mu_c \right)^{- \frac{1}{2}} + \mathcal{O}\left(  \left( \mu-\mu_c \right)^{\frac{1}{2}} \right)~.
\end{align}
Thus, the second derivative of the defect free energy diverges approaching $\mu_c$ from the right, with critical exponent $\frac{1}{2}$.
Integrating \eqref{eq:dsquareFDYN} twice and imposing initial and junction condition leads to 
\begin{align}\label{eq:FS3-YN}
	\frac{F_{\mathcal{D}}^{Y_N}}{6\pi (1-\Delta)N^2} &=\begin{cases}
		- \mu~, & \mu< - \frac{\ln 2}{2\pi} \\
		-\mu+\frac{2}{\pi^2}\Im\Li_2\left(\frac{1+i \sqrt{4 e^{4 \pi  \mu }-1}}{2}\right) + \frac{4}{\pi}\mu \tan^{-1}\left(\sqrt{4e^{4\pi \mu}-1}\right)~, & \mu>-\frac{\ln 2}{2\pi} ~.
	\end{cases}
\end{align}
Noting that $\Li_2(z)$ and $\Li_2(z/(z-1))$ differ only by log terms, the dilogarithm term in the above result equals the one in \eqref{eq:massiveSD3YNsugra}. So we once again find a match between the field theory and string theory results, including a match of the phase transition.

The $Y_N$ defect phase transition bears resemblance with a transition observed in 3d SQED \cite{Russo:2016ueu,Santilli:2019mtc}. Both are second order, controlled by the real mass parameter, and with critical exponent~$\frac{1}{2}$. We note that the defect phase transition in $Y_N$ is in a different universality class than the more customary third order ones with finite discontinuity that appear in the planar limit \cite{Nedelin:2015mta,Santilli:2021qyt}.\footnote{A phase transition controlled by a mass parameter associated with a probe D7-brane in AdS$_5$/CFT$_4$ was discussed in \cite{Karch:2015kfa}. The critical exponents found here differ from those of \cite{Karch:2015kfa} as well.}

\bigskip
\textbf{Defect $F$-theorem:} Our results for the mass deformed gauge theory defects are consistent with the defect $F$-theorem \cite{Casini:2023kyj}. The mass deformation triggers an RG flow from the conformal defect to an empty defect. At large $\mu$, $F_{\cD}$ grows linearly, and matches the free energy of a 3d chiral/antichiral pair of scaling dimension $\Delta$ in the fundamental representation of $SU(N_{t_0})$: 
\begin{align}\label{eq:largemuFD}
	\ && F_{\cD} \approx 2 \pi (1-\Delta) N_{t_0} \lvert m \rvert , && m= 3L \mu , \ \mu \to \pm \infty ~.
\end{align}
The linear part of $F_{\cD}$ is scheme-dependent \cite{Klebanov:2011gs,Closset:2012vg}. We choose a scheme that removes the linear divergence \eqref{eq:largemuFD}, exactly as for a standalone 3d massive chiral/antichiral pair.
The resulting $F_{\cD}$ is positive, decreases monotonically, and vanishes at $\lvert \mu \rvert \to \infty$.

\section{Connection to M-theory} \label{sec:Mtheory}

\begin{figure}
	\centering
	\subfigure[][]{\label{fig:toricdiags-1}
		\includegraphics[height=0.15\textwidth]{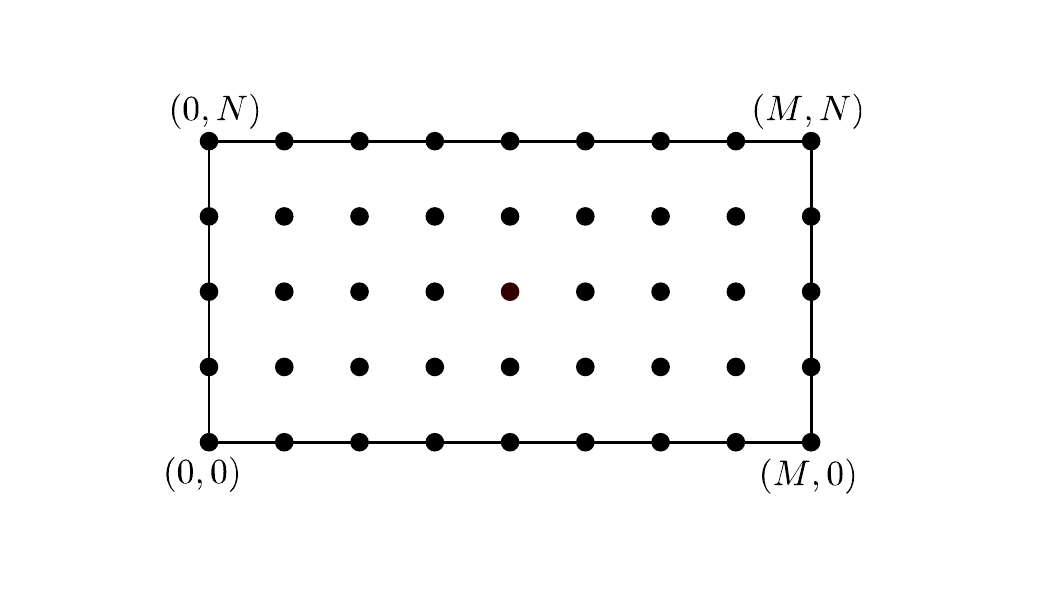}
	}\hskip 10mm
	\subfigure[][]{\label{fig:toricdiags-2}
		\includegraphics[height=0.15\textwidth]{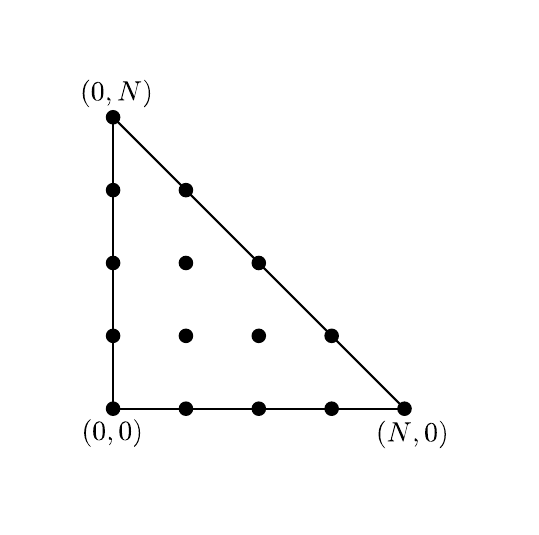}
	}\hskip 10mm
	\subfigure[][]{\label{fig:toricdiags-3}
		\includegraphics[height=0.15\textwidth]{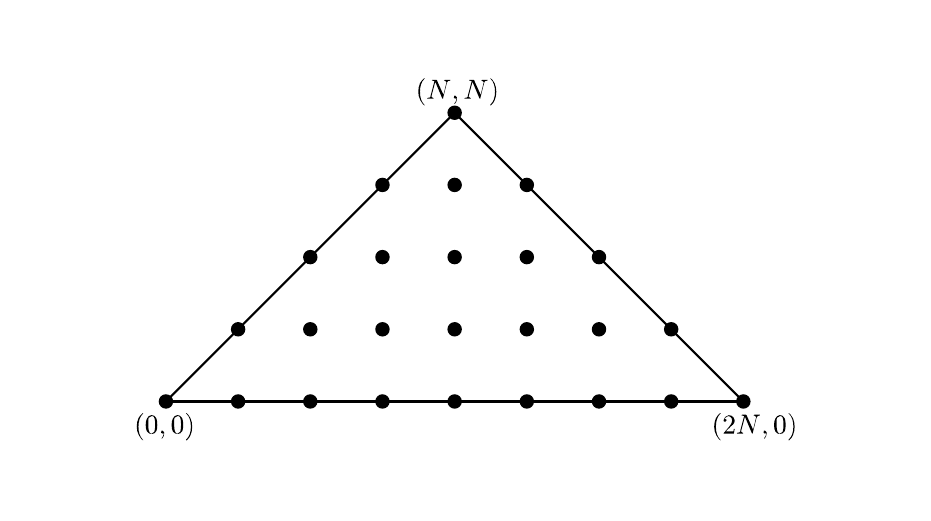}
	}
	\caption{Toric polygons describing the Calabi--Yau threefold singularities corresponding to the 5d SCFTs engineered by the brane webs in fig.~\ref{fig:pqwebs}: (a) $+_{N,M}$, (b) $T_N$ and (c) $Y_N$.}
	\label{fig:toricdiags}
\end{figure}

The codimension-2 defects we consider can also be geometrically engineered in M-theory.
On the one hand, the 5d SCFTs of interest can be realized from M-theory compactifications on a toric, local Calabi--Yau threefold singularity $X$ \cite{Intriligator:1997pq,Xie:2017pfl,Closset:2018bjz,Apruzzi:2019opn,Apruzzi:2019enx,Eckhard:2020jyr}. The toric polygons engineering the three sample SCFTs of fig.~\ref{fig:pqwebs} are in fig.~\ref{fig:toricdiags}. 
We now discuss how to realize the defects of interest in M-theory, building on \cite{Dimofte:2010tz,Banerjee:2018syt,Banerjee:2019apt,Banerjee:2020moh}.
We then consider the mirror geometry $X^{\vee}$ to $X$, which engineers the same 5d SCFT, and construct codimension-2 defects also from this mirror perspective.

\par

\begin{figure}
	\centering
	\subfigure[][]{\label{fig:toricdresol-1}
		\begin{tikzpicture}[scale=0.55]
			\draw[blue] (0,0) -- (1,0) -- (1,1) -- (0,1) -- (0,0);
			\draw[blue] (0,0) -- (-0.7,-0.7) -- (-1.7,-0.7);
			\draw[blue] (-0.7,-0.7) -- (-0.7,-1.7);
			\draw[blue] (1,0) -- (1.7,-0.7) -- (2.7,-0.7);
			\draw[blue] (1.7,-0.7) -- (1.7,-1.7);
			\draw[blue] (0,1) -- (-0.7,1.7) -- (-1.7,1.7);
			\draw[blue] (-0.7,1.7) -- (-0.7,2.7);
			\draw[blue] (1,1) -- (1.7,1.7) -- (2.7,1.7);
			\draw[blue] (1.7,1.7) -- (1.7,2.7);
			
			\node at (-1.2,0.5) {$\bullet$};
			\node at (0.5,0.5) {$\bullet$};
			\node at (2.2,0.5) {$\bullet$};
			\node at (-1.2,2.2) {$\bullet$};
			\node at (0.5,2.2) {$\bullet$};
			\node at (2.2,2.2) {$\bullet$};
			\node at (-1.2,-1.2) {$\bullet$};
			\node at (0.5,-1.2) {$\bullet$};
			\node at (2.2,-1.2) {$\bullet$};
			
			\draw[black,thick] (-1.2,-1.2) -- (2.2,-1.2) -- (2.2,2.2) -- (-1.2,2.2) -- (-1.2,-1.2);
			\draw[black,thick] (-1.2,0.5) -- (0.5,-1.2) -- (2.2,0.5) -- (0.5,2.2) -- (-1.2,0.5);
			\draw[black,thick] (-1.2,0.5) -- (2.2,0.5);
			\draw[black,thick] (0.5,-1.2) -- (0.5,2.2);
			
		\end{tikzpicture}
	}\hskip 20mm
	\subfigure[][]{\label{fig:toricresol-2}
	\begin{tikzpicture}
			
			\node at (-1.8,-1.8) {$\bullet$};
			\node at (0,0) {$\bullet$};
			\node at (-1.8,0) {$\bullet$};
			\node at (0,-1.8) {$\bullet$};
			\draw[black,thick] (-1.8,0) -- (0,0) -- (0,-1.8) -- (-1.8,-1.8) -- (-1.8,0) -- (0,-1.8);
			
			\draw[blue] (-2.2,-1.2) -- (-1.2,-1.2) -- (-0.6,-0.6) -- (0.2,-0.6);
			\draw[blue] (-1.2,-1.2) -- (-1.2,-2.2);
			\draw[blue] (-0.6,-0.6) -- (-0.6,0.2);
			
			\node[] at (-1.2,-1.2) {\begin{color}{gray}$\blacklozenge$\end{color}};
			
		\end{tikzpicture}
	}\hskip 10mm
	
	\caption{Left: Resolved 5-brane web for $+_{2,2}$ (blue) and its dual toric polygon (black). Right: Resolution of the brane junction where the D3-brane, indicated by \begin{color}{gray}$\blacklozenge$\end{color}, is located. The junction can be embedded in a larger 5-brane web. $\mathfrak{L}$ is specified by the point \begin{color}{gray}$\blacklozenge$\end{color} in the face of the polygon.}
	\label{fig:toricresol}
\end{figure}
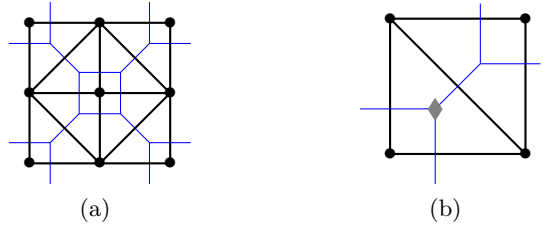

\textbf{Lagrangian defects:} Let $\widetilde{X} \to X$ be a crepant resolution of the toric Calabi--Yau singularity $X$, i.e. $\widetilde{X}$ is a smooth toric Calabi--Yau encoded in a triangulation of the toric polygons of fig.~\ref{fig:toricdiags}. There is a direct connection between 5-brane webs and M-theory on $\widetilde{X}$ \cite{Leung:1997tw}: the resolved $(p,q)$ 5-brane web is the dual graph of the triangulated toric polygon of $\widetilde{X}$, see fig.~\ref{fig:toricdresol-1} for an example.
M-theory on $X \times \RR^{1,4}$ gives rise to a 5d SCFT, and codimension-2 half-BPS defects arise from M5-branes wrapping special Lagrangian three-cycles $\mathfrak{L} \subset X$ and extending in spacetime along $\RR^{1,2} \subset \RR^{1,4}$:
\begin{equation}\label{eq:M5-cL}
	\RR^{1,2} \times \mathfrak{L} \  \subset \ \RR^{1,4} \times X ~.
\end{equation}
Defects in the 5d gauge theory are engineered by special Lagrangians in the resolved Calabi--Yau, $\mathfrak{L} \subset \widetilde{X}$. The choice of $\mathfrak{L}$ we need is locally an $S^1$-fibration over $\RR^2$ \cite{Dimofte:2010tz,Banerjee:2018syt}.

A Lagrangian in $\widetilde{X}$ is represented by a line transverse to the plane of the toric polygon and piercing it at a point, see for example \cite{Aganagic:2000gs,Aganagic:2001nx}. 
Therefore, $\mathfrak{L}$ is specified by a face of the toric polygon of $\widetilde{X}$ (fig.~\ref{fig:toricresol-2}). The semi-classical moduli space $\cM_{\mathfrak{L}}$ of special Lagrangians is two-real dimensional, and the two real coordinates on $\cM_{\mathfrak{L}}$ are identified with horizontal and vertical motion in the toric diagram. This geometric prescription to engineer a codimension-2 defect is dual to the D3-brane prescription, as exemplified in fig.~\ref{fig:toricresol-2}.

We introduce the complexification $\cM_{\mathfrak{L}}^{\CC}$ of $\cM_{\mathfrak{L}}$ and denote points on it by the pair of complex numbers $(c_1,c_2)$, with the convention that $\Re (c_1)=0=\Re (c_2)$ is the fixed point of the isometries of the toric polygon of $X$. This point gives us a reference Lagrangian $\mathfrak{L}$, and the other elements of $\cM_{\mathfrak{L}}$ are geometric deformations of it.

\par

\bigskip
\textbf{Mirror:} The Hori--Vafa mirror $X^{\vee}$ of $X$ is a hypersurface 
\begin{align}
	X^{\vee} = \left\{ (s,t,u,v) \in \CC^4  \ : \ u v = P(s,t)	\right\} ~,
\end{align}
with a polynomial $P (s,t)$ read off from the toric polygon of $X$. Explicitly, 
\begin{align}
	P (s,t) = \sum_{(k,\ell)} \xi_{k \ell } s^k t^{\ell}~,
\end{align}
where the sum runs over lattice points $(k,\ell) \in \ZZ^2$ that belong to the toric polygon of $X$, and $\xi_{k\ell}$ are deformation parameters. By redefinitions of $(s,t,u,v)$, we can set $\xi_{00}=1$ and also set to $- 1$ (resp. $+1$) the coefficients of the monomial of highest degree in $s$ (resp. $t$). The remaining $\xi_{k\ell}$ have the meaning of Higgs branch parameters, and the SCFT is recovered by setting them to zero.

Therefore, the properties of the SCFT are encoded in the mirror curve 
\begin{align}\label{MirrorCurve}
	\Sigma_{\rm M5} \ : \ &&  P(s,t) = 0~, &&
\end{align}
studied e.g.\ in \cite{Kol:1997fv}. An explicit mapping between $\Sigma_{\rm M5} $ and the surface $\Sigma$ appearing in the $\mathrm{AdS}_6 \times S^2 \times \Sigma$ supergravity solutions of \cite{DHoker:2016ujz,DHoker:2016ysh,DHoker:2017mds,DHoker:2017zwj} was given in \cite{Kaidi:2018zkx}. 
The functions $\cA_{\pm}$ of sec.~\ref{sec:holography} can be written as \cite{Kaidi:2018zkx}
\begin{align}\label{eq:cApmtoM5}
	\cA_{\pm} = \frac{3}{8\pi} \left( i N^2\ln s \pm L^2 \ln t \right)~,
\end{align}
where $N$ and $L$ are the degree of $P(s,t)$ in $s$ and $t$, respectively.\footnote{Our conventions differ slightly from \cite{Kaidi:2018zkx}, so that $P(s,t)$ is defined as in \cite{Aganagic:2000gs}.}
Through this relation, the Type IIB $\rm AdS_6$ solutions provide an explicit description of $ \Sigma_{\rm M5}$: 
By solving (\ref{eq:cApmtoM5}) for $s$, $t$ in terms of $\cA_\pm$, which are in turn expressed in terms of a coordinate $w$ on $\Sigma$, $w$ parametrizes a solution to \eqref{MirrorCurve}, 
\begin{align}\label{eq:wparamst}
	w \mapsto (s(w),t(w)) \in \Sigma_{\rm M5}~.
\end{align}

The mirror of the special Lagrangian $\mathfrak{L}$ is determined as follows \cite{Aganagic:2000gs}: We start with the complexified moduli space $\cM_{\mathfrak{L}}^{\CC}$. At every point $(c_1, c_2)$ on it, $(\Im (c_1), \Im (c_2))$ are fixed by the complex equation \eqref{MirrorCurve}. This leaves two real parameters, that identify a point $\mathfrak{L}$ in the moduli space $\cM_{\mathfrak{L}}$ of special Lagrangian defects.
By mirror symmetry, $\mathfrak{L}$ is mapped to a holomorphic cycle in $X^{\vee}$ located at a point $(s(c_1,c_2), t(c_1,c_2))$ on the mirror curve. The topology of $\mathfrak{L}$ we need is $S^1 \times \RR^2$, or a compactification to $S^3$. In this case, the image of $(c_1,c_2)$ in $\Sigma_{\rm M5}$ takes the simpler from $(s(c_1), t(c_2))$ \cite{Aganagic:2000gs,Aganagic:2001nx}. We invert the relation and write the complex numbers $(c_1,c_2)$ as functions of the coordinates $(s,t)$.\footnote{When taking the mirror of $\mathfrak{L}$, one should account for a frame factor \cite{Aganagic:2001nx}. Here we fix it to match with \cite{Kaidi:2018zkx}.} By virtue of \eqref{eq:wparamst}, $(c_1,c_2)$ are parametrized by the complex coordinate $w$ on $\Sigma_{\rm M5}$.

We conclude that the choice of element in the moduli space of Lagrangian defects in M-theory is uniquely determined by a point on $\Sigma$. This fact matches the Type IIB construction of \cite{Gutperle:2020rty} and the discussion in sec.~\ref{sec:holography}.

\bigskip
\textbf{Reduction to Type IIA:} Starting with an M5-brane wrapping \eqref{eq:M5-cL}, we can compactify the direction $x^2$ and interpret it as the M-theory circle. Reducing along that direction leads to Type IIA string theory with a D4-brane wrapping $\mathfrak{L}$ and extending along $\RR^{1,1} \subset \RR^{1,3}$. This configuration engineers a codimension-2 defect in a 4d gauge theory.

At this point we can apply T-duality along three distinct directions inside $X$, letting it act twice inside $\mathfrak{L}$, and land on $X^{\vee}$ \cite{Strominger:1996it}. We end up with a D3-brane wrapping a holomorphic two-cycle in $X^{\vee}$ and located at a point on $\Sigma_{\rm M5}$, extending along $\RR^{1,1}$. The same situation is recovered by the direct dimensional reduction from our D3-brane defect configuration. This result is also consistent with the realization of the defect directly in a 4d theory of class S characterized by the curve $\Sigma_{\rm M5}$.

\subsection{Lagrangian defect case studies}
We now exemplify the construction of 3d planar defects along $\RR^{1,2}$ from M5-branes on a special Lagrangian $\mathfrak{L}  \cong S^1 \times \RR^2$. We discuss the toric Calabi--Yau examples of fig.~\ref{fig:toricdiags}. We work directly with the singularity $X$, rather than with a resolution $\widetilde{X}$, thus focusing on the case in which the ambient 5d theory is conformal.

\bigskip
\textbf{\boldmath{$T_{N}$} theory:} The mirror curve for the polygon in fig.~\ref{fig:toricdiags-2} is 
\begin{align}\label{eq:SigmaM5TN}
	\Sigma_{\rm M5}^{T_N} \ : \ &&  t^{N}-s^{N} +1= 0~, &&
\end{align}
where we are turning off the deformations of the mirror geometry. The relation between $(s,t)$ and the embedding coordinate $w$ obtained from (\ref{eq:cApmtoM5}) and (\ref{eq:cApmTN}) is \cite{Kaidi:2018zkx} 
\begin{align}\label{eq:TNstfromw}
	s= \left(\frac{2w}{w+1}\right)^{1/N} ~, && t = \left(\frac{w-1}{w+1}\right)^{1/N} ~.
\end{align}
To construct a Lagrangian defect, we start with the somewhat degenerate case $N=1$, corresponding to the smooth $X= \CC^3$, and generalize afterwards. A Lagrangian with topology $S^1 \times \RR^2$ in $\CC^3$ is determined by two charge vectors $q^{a}= (q_{j} ^{a})_{j=1,2,3}$ satisfying 
\begin{equation}
	\sum_{j=1}^{3} q_j ^{a} =0~,
\end{equation}
for each $a=1,2$. The mirror of this Lagrangian is \cite{Aganagic:2000gs}
\begin{align}\label{eq:tstoc1c2TN}
	t = e^{2c_1}~, &&  s= e^{2c_2} ~.
\end{align}
The construction generalizes to $T_N$. Plugging \eqref{eq:tstoc1c2TN} into \eqref{eq:SigmaM5TN} fixes $\Im (c_1)$ and $\Im (c_2)$ in terms of the real parts.
Using this dictionary and comparing with \cite[(3.17)]{Gutperle:2020rty}, we identify the relation 
\begin{align}\label{eq:relm1m2c1c2TN}
	\Re (c_1)= m_1e^{-r} ~, && \Re (c_2)= m_2e^{-r}
\end{align}
between the geometric parameters $(\Re (c_1), \Re (c_2))$ and the mass parameters $(m_1,m_2)$.
Furthermore, from \eqref{eq:TNstfromw}, deformations of the defect are parametrized by the local complex coordinate $w$. 
For the conformal defect with $w_c= i/\sqrt{3}$,
\begin{align}
	t = e^{2\pi i/3} ~, && -s=e^{4 \pi i/3} ~.
\end{align}
This yields the fixed point of the $\ZZ_3$ symmetry permuting $t^N, -s^N, 1$ in \eqref{eq:SigmaM5TN}. Undoing the mirror map, $w_c$ is sent to $\Re c_1 = 0 = \Re c_2$, recovering the undeformed Lagrangian defect in M-theory, as expected.\par

\bigskip
\textbf{\boldmath{$Y_{N}$} theory:} The mirror curve for the polygon in fig.~\ref{fig:toricdiags-3} is
\begin{align}
	\Sigma_{\rm M5}^{Y_N} \ : \ && (st)^{N}-s^{2N} +1= 0~. &&
\end{align}
With a change of coordinates, it is equivalent to \eqref{eq:SigmaM5TN}. 
The 3d defect geometrically engineered by $\mathfrak{L}$ is therefore analogous to the $T_N$ case, in agreement with the Type IIB analysis  \cite{Gutperle:2020rty}.

\bigskip
\textbf{\boldmath{$+_{N,M}$} theory:}  As last example, we discuss $+_{N,M}$ as ambient theory. The mirror curve to the Calabi--Yau singularity of fig.~\ref{fig:toricdiags-1} is 
\begin{align}\label{eq:curveplusNM}
	\Sigma_{\rm M5}^{+_{N,M}} \ : \ && s^{N} + t^M-s^{N}t^M +1= 0~. &&
\end{align}
This curve can be parametrized as $w \mapsto (s(w),t(w))$ with \cite{Kaidi:2018zkx}
\begin{align}
	s= \left(\frac{2w-1}{w-1}\right)^{1/N} ~, && t = \left(\frac{3w-2}{w}\right)^{1/M} ~.
\end{align}
For the special case $+_{1,1}$, $\mathfrak{L}$ was given in \cite{Aganagic:2000gs}, and its construction is analogous to the previous examples. We employ it in the $+_{N,M}$ theory and, substituting for the location of the mirror of $\mathfrak{L}$, we find again relation \eqref{eq:relm1m2c1c2TN} between $(\Re (c_1), \Re (c_2))$ and $(m_1,m_2)$. The polygon in fig.~\ref{fig:toricdiags-1} possesses a $\ZZ_2 \times \ZZ_2$ symmetry, which is inherited by \eqref{eq:curveplusNM}, with the two $\ZZ_2$ factors acting as 
\begin{align}
	s^k \mapsto s^{N-k} ~, && t^{\ell} \mapsto t^{M-\ell}~,
\end{align}
for $k=0, \dots, N$ and $\ell=0, \dots, M$. The conformal embedding $w_c$ corresponds to the fixed point of this symmetry.

As a consistency check, we can reduce the toric Calabi--Yau with an M5-brane wrapping $\mathfrak{L}$ to Type IIA. Applying T-duality twice leads to a D2-brane ending on the intersection of $N$ D4 and an $M$ NS5-branes. This setup generalizes \cite[sec.~3.2.2]{Dimofte:2010tz} and describes a 2d $\mathcal{N}=(2,2)$ defect in a 4d SCFT. It agrees with the result of applying T-duality to the D3-brane defect in the $+_{N,M}$ 5-brane web.

\let\oldaddcontentsline\addcontentsline
\renewcommand{\addcontentsline}[3]{}
\begin{acknowledgments}
	We thank Julius Grimminger, Hee-Cheol Kim, Mauricio Romo and Lorenzo Ruggeri for discussion.
	CFU was supported, in part, by the US Department of Energy under Grant No.~DE-SC0007859 and by the Leinweber Center for Theoretical Physics. LS was supported by the Funda\c{c}\~{a}o para a Ci\^{e}ncia e a Tecnologia project PTDC/MAT-PUR/30234/2017 during the early stages of the work.
	This work was initiated at the Aspen Center for Physics, which is supported by National Science Foundation grant PHY-1607611.
\end{acknowledgments}
\let\addcontentsline\oldaddcontentsline

\appendix
\renewcommand\theequation{\thesection.\arabic{equation}}

\section{D3-brane BPS conditions}\label{app:BPS}

We derive the BPS conditions, integrate them, and integrate the on-shell action for defects along $\rm AdS_3$ and $S^3$.


\bigskip
\textbf{Defect RG flows on \boldmath{$\RR^{1,2}\subset \RR^{1,4}$}:}
For a 5d SCFT on $\RR^{1,4}$ the $\rm AdS_6$ metric is taken as \eqref{eq:AdS6-metric-flat}.
The BPS condition for D3-branes wrapping $\RR^{1,2}$ and location on $\Sigma$ depending on the $\rm AdS_6$ radial coordinate, parametrized by $w(r)$, was derived in \cite{Gutperle:2020rty} and reads
\begin{equation}
	\label{eq:BPSeqkappa}
	\kappa^2 w^{\prime} = \partial_{\bar{w}} \mathcal{G}~.
\end{equation}
It is solved by
\begin{equation}
	\label{eq:BPSAandM}
	\cA_{+} (w) - \overline{\cA}_{-} (\bar w) = m e^{-r}~,
\end{equation}
The general D3-brane action for such an embedding is
\begin{equation}
	\label{eq:SD31}
	S_{\rm D3} = T_{\rm D3} \Vol_{\RR^{1,2}} \int dr\, e^{3r} f_6 ^4 \sqrt{ 1+ 4 \frac{ \rho^2 \lvert w^{\prime} \rvert^2 }{f_6 ^2}}~,
\end{equation}
where $w^{\prime} \equiv \partial_r w$. 
Following \cite[app.~A]{Gutperle:2020rty}, the action for BPS solutions becomes 
\begin{align}
	S_{\rm D3} = T_{\rm D3} \Vol_{\RR^{1,2}} \int dr\, e^{3r}\,6 \cG T^2 ~.
\end{align}
Using the definition of $T^2$ from \eqref{eq:kappa2-G} and the BPS condition, we may write this as
\begin{align}
	S_{\rm D3} = T_{\rm D3} \Vol_{\RR^{1,2}} \int dr\, e^{3r}\,2\left(3\cG+w^\prime\partial_{w}\cG +\bar{w}^\prime\partial_{\bar{w}}\cG \right)
	=2T_{\rm D3} \Vol_{\RR^{1,2}} \int dr\, \frac{d \ }{dr}\left[e^{3r}\cG\right]~.
\end{align}

\bigskip
\textbf{Defect RG flows on \boldmath{$\rm AdS_3\subset AdS_5$}:}
We now turn to D3-brane embeddings in the geometry \eqref{eq:metric-AdS6-AdS5}.
The D3 action then is
\begin{align}\label{d3act-ads}
	S_{\rm D3}&=T_{\rm D3}\Vol_{\rm AdS_3}\int dr\, f_6^3 \cosh^3\!r\,\sqrt{f_6^2+4\rho^2 |w^\prime|^2}~.
\end{align}
With the explicit form of the metric functions the Lagrangian becomes
\begin{align}\label{d3-L-ads}
	L_{\rm D3}&=6\cosh^3\!r\,\cG T \sqrt{1+\frac{2\kappa^2}{3\cG}|w^\prime|^2}~.
\end{align}

The BPS solutions can be determined from the requirement that the square root in \eqref{d3-L-ads} has to uniformize for the BPS conditions to be solvable. Taking inspiration from the BPS condition for planar defects in (\ref{eq:BPSeqkappa}), we make the ansatz
\begin{align}\label{eq:BPS-mass-ads-0}
	\kappa^2 w^\prime &= e^{i f(r)}\partial_{\bar w}\cG~,
\end{align}
with a real function $f$. The extremality condition for the embedding with this ansatz becomes
\begin{align}
	\partial_r \frac{\delta L_{\rm D3}}{\delta \bar w^\prime}-\frac{\delta L_{\rm D3}}{\delta \bar w}
	&=2\partial_{\bar w}\cG \cosh^3\!r\,\left[i f' e^{if}+3e^{if}\tanh r
	-e^{2if}-2\right]=0~.
\end{align}
This is one complex equation for one real function. So this is overconstrained, as one may expect for such an ansatz. Nevertheless, we find a (real) solution
\begin{align}\label{eq:f-ads}
	f&=\cos^{-1}\left(\tanh r\right)~.
\end{align}
This leads to the BPS equation
\begin{align}\label{eq:BPS-mass-ads}
	\kappa^2 w^\prime &= \frac{e^r+i}{e^r-i}\,\partial_{\bar w}\cG~.
\end{align}

To integrate the BPS condition we spell it out more explicitly,
\begin{align}
	0&=\partial_{\bar{w}}\overline{\cA}_+\left(\cA_+-\overline{\cA}_-+e^{-if}w'\partial_w\cA_+ \right)+
	\partial_{\bar{w}}\overline{\cA}_-\left(\overline{\cA}_+-\cA_--e^{-if}w'\partial_w\cA_- \right)~.
\end{align}
By adding zero we can turn the terms in the brackets into complex conjugates of each other
\begin{align}
	0=\,&\partial_{\bar{w}}\overline{\cA}_+\left(\cA_+-\overline{\cA}_-+e^{-if}w'\partial_w\cA_+-e^{if}\bar w'\,\partial_{\bar{w}}\overline{\cA}_- \right)
	\nonumber\\&+
	\partial_{\bar{w}}\overline{\cA}_-\left(\overline{\cA}_+-\cA_--e^{-if}w'\partial_w\cA_- +e^{if}\bar w'\,\partial_{\bar{w}}\overline{\cA}_+\right)~.
\end{align}
The BPS condition is therefore solved by 
\begin{align}
	\cA_+-\overline{\cA}_-+e^{-if} \frac{d}{dr}\cA_+ -e^{if}\frac{d}{dr}\cA_- &=0~,
\end{align}
where $e^{if}=(e^r+i)/(e^r-i)$. We may integrate this to 
\begin{align}
	\sech r\frac{d}{dr}\left[(\sinh r-i)\cA_+-(\sinh r+i)\overline{\cA}_-\right]&=0
\end{align}
This leads to the BPS condition in \eqref{eq:BPS-AdS}.

We now evaluate the action on BPS embeddings. The first step is not affected by the phase in \eqref{eq:BPS-mass-ads} compared to the planar defect, and we find
\begin{align}
	S_{\rm D3} = T_{\rm D3} \Vol_{\rm AdS_3} \int dr\, \cosh^3\!r\,6 \cG T^2 ~.
\end{align}
Using again the definition of $T^2$ and the BPS condition leads to 
\begin{align}
	S_{\rm D3} = 2T_{\rm D3} \Vol_{\rm AdS_3} \int dr\, \cosh^3\!r\,\left[3\cG+e^{-if}w'\partial_w\cG+e^{if}\bar w'\,\partial_{\bar{w}}\cG\right] ~.
\end{align}
We can write this as
\begin{align}
	S_{\rm D3} = 2T_{\rm D3} \Vol_{\rm AdS_3} \int dr\,
	\left(\frac{d}{dr}(h(r) \cG)-
	2\sinh r\left(w'\partial_w\cG+\bar w'\,\partial_{\bar{w}}\cG\right)
	-i\cosh^2\!r\,\left(w'\partial_w\cG-\bar w'\,\partial_{\bar{w}}\cG\right)
	\right)~,
\end{align}
where $h(r)=\frac{1}{4}(\sinh(3r)+9\sinh(r))$ satisfies $h'(r)=3\cosh^3\!r$.
From the BPS condition in \eqref{eq:BPS-mass-ads} we conclude
\begin{align}
	\kappa^2 \lvert w' \rvert^2 &=e^{if}\partial_{\bar{w}}\cG\bar{w}^\prime
	&&\Rightarrow&
	\Im\left(e^{if}\partial_{\bar{w}}\cG\bar{w}^\prime\right)&=0~.	
\end{align}
The second equality follows since the quantity on the left hand side of the first equation is real. With the explicit solution for $f$,
\begin{align}
	-i\sinh r\left(w'\partial_w\cG-\bar{w}^\prime\partial_{\bar{w}}\cG\right)&=w'\partial_w\cG+\bar{w}^\prime\partial_{\bar{w}}\cG~.
\end{align}
This leads to
\begin{align}
	S_{\rm D3} &= 2T_{\rm D3} \Vol_{\rm AdS_3} \int dr\,
	\left(\frac{d}{dr}(h(r) \cG)-
	\sinh r\left(w'\partial_w\cG+\bar w'\,\partial_{\bar{w}}\cG\right)
	-i\left(w'\partial_w\cG-\bar w'\,\partial_{\bar{w}}\cG\right)
	\right)~.
\end{align}
From the BPS condition,
\begin{align}
	\sinh r\, w'\partial_w\cG&=\sinh r\left(\overline{\cA}_+-\cA_-\right)\partial_w\cA_+w'
	+\sinh r\left(\cA_+-\overline{\cA}_-\right)\partial_w\cA_-w'
	\nonumber\\
	&=\frac{1}{2}\left(\bar m\partial_w\cA_++m\partial_w\cA_-\right)w'
	-i\left(\overline{\cA}_++\cA_-\right)\partial_w\cA_+w'
	+i\left(\cA_++\overline{\cA}_-\right)\partial_w\cA_-w'
	\nonumber\\
	&=\frac{d}{dr}\left[\frac{1}{2}(\overline{m}\cA_++m\cA_-)\right]+iw'\partial_w\cB - i \overline{\cA}_+\partial_w \cA_+ w'
	+i \overline{\cA}_-\partial_w \cA_- w'
\end{align}
Noting that $\partial_w\cG=\overline{\cA}_+\partial_w\cA_+-\overline{\cA}_-\partial_w\cA_-+\partial_w \cB$, we arrive at
\begin{align}
	\sinh r\, w'\partial_w\cG+iw'\partial_w\cG&=
	\frac{d}{dr}\left[
	\frac{1}{2}(\overline{m}\cA_++m\cA_-)+2i\cB
	\right].
\end{align}
In summary,
\begin{align}
	S_{\rm D3} &= 2T_{\rm D3} \Vol_{\rm AdS_3} \int dr\,
	\frac{d}{dr}\left[h(r) \cG
	-\frac{1}{2}(\overline{m}\cA_++m\cA_-+m\overline{\cA}_++\overline{m}\overline{\cA}_-)+2i\cB-2i\overline{\cB}
	\right].	
\end{align}

\bigskip
\textbf{Defect RG flows on \boldmath{$S^3\subset S^5$}:}
We now turn to the geometry in \eqref{eq:AdS6-metric-S5}. 
After the analytic continuation $r\rightarrow r+\frac{i\pi}{2}$ detailed above (\ref{eq:AdS6-metric-S5}), the BPS conditions are
\begin{align}\label{eq:BPS-cApm-S}
	(\cosh r-1)\cA_+-(\cosh r+1)\overline{\cA}_-&=-\frac{im}{2}~,
	\nonumber\\
	(\cosh r+1)\overline{\cA}_+-(\cosh r-1)\cA_-&=-\frac{i\overline{m}}{2}~,
\end{align}

The first step in evaluating the action is again not affected by the extra coefficient in \eqref{eq:BPS-mass-ads} compared to the planar defect, even if it is not a phase anymore after analytic continuation. We find
\begin{align}
	S_{\rm D3} = T_{\rm D3} \Vol_{\rm S^3} \int dr\, \sinh^3\!r\,6 \cG T^2 ~.
\end{align}
Using again the definition of $T^2$ and the BPS condition leads to 
\begin{align}
	S_{\rm D3} = 2T_{\rm D3} \Vol_{\rm S^3} \int dr\, \sinh^3\!r\,\left[3\cG+e^{-if}w'\partial_w\cG+e^{if}\bar w'\,\partial_{\bar{w}}\cG\right] 
\end{align}
with $e^{if}=\coth(r/2)$ and $e^{-if}=\tanh(r/2)$.
We can write this as
\begin{align}
	S_{\rm D3} = 2T_{\rm D3} \Vol_{\rm S^3} \int dr\,
	\left[\frac{d}{dr}(\tilde{h}(r) \cG)+
	2\cosh r\left(w'\partial_w\cG+\bar w'\,\partial_{\bar{w}}\cG\right)
	-\sinh^2\!r\,\left(w'\partial_w\cG-\bar w'\,\partial_{\bar{w}}\cG\right)
	\right],
\end{align}
where $\tilde{h}(r)=\frac{1}{4}(\cosh(3r)-9\cosh(r))$ satisfies $\tilde{h}'(r)=3\sinh^3\!r$.
As for the $\rm AdS_3$ defects before, we derive from the BPS condition
\begin{align}
	0&=\kappa^2 w^\prime\bar{w}^\prime-\kappa^2 w^\prime\bar{w}^\prime = \bar{w}^\prime e^{if}\partial_{\bar{w}}\cG-w^\prime e^{-if}\partial_w\cG
	\nonumber\\
	&=\frac{e^{if}-e^{-if}}{2}\left(\bar{w}^\prime\,\partial_{\bar{w}}\cG+w^\prime\partial_w\cG\right)
	+\frac{e^{if}+e^{-if}}{2}\left(\bar{w}^\prime\,\partial_{\bar{w}}\cG-w^\prime\partial_w\cG\right)
	\nonumber\\
	&=\csch r\left[\bar{w}^\prime\,\partial_{\bar{w}}\cG+w^\prime\partial_w\cG
	+\cosh r\,\left(	\bar{w}^\prime\,\partial_{\bar{w}}\cG-w^\prime\partial_w\cG\right)\right].
\end{align}
This leads to
\begin{align}
	S_{\rm D3} = 2T_{\rm D3} \Vol_{\rm S^3} \int dr\,
	\left[\frac{d}{dr}(\tilde{h}(r) \cG)+
	\cosh r\left(w'\partial_w\cG+\bar w'\,\partial_{\bar{w}}\cG\right)
	+w'\partial_w\cG-\bar w'\,\partial_{\bar{w}}\cG\right].
\end{align}
With the integrated BPS condition,
\begin{align}
	S_{\rm D3} = 2T_{\rm D3} \Vol_{\rm S^3} \int dr\,
	\frac{d}{dr}\left[ h(r) \cG+2\cB-2\overline{\cB}
	-\frac{i}{2}\left(\overline{m}\cA_++m\cA_-+m\overline{\cA}_++\overline{m}\overline{\cA}_-\right)
	\right].
\end{align}

\section{Codimension-2 defects in 5d \texorpdfstring{$\cN=2$}{N=2} SYM}
\label{app:MSYM}

An immediate application of our localization computation in sec.~\ref{sec:localization} is to study codimension-2 defects in maximally supersymmetric 5d $SU(N)$ Yang--Mills theory ($\cN=2$ SYM). This gauge theory is not conformal, and is a low energy description of the 6d $\mathcal{N}=(2,0)$ $\mathfrak{su}(N)$ CFT compactified on a $S^1$ of radius $r_6$. The precise relation between $r_6$ and the 5d gauge coupling $g_{\text{\tiny YM}}^2$ is \cite{Kallen:2012zn}
\begin{align}
	r_6 ^{-1} =  \frac{8 \pi^2}{g_{\text{\tiny YM}}^2} ~.
\end{align}
The perturbative partition function on $S^5$ of radius $r_5$ is 
\begin{equation}
	\cZ_{S^5} \left[ \cD \right] = \left[ \prod_{j=1}^{N} \int_{-\infty}^{\infty} d \lambda_j \right] \delta \left( \sum_{j=1}^N \lambda_j \right) \prod_{1\le j<k \le N} [2 \sinh \pi (\lambda_j - \lambda_k)]^2 \prod_{j=1}^{N} \frac{e^{- \frac{8 \pi^3 r_5}{g_{\text{\tiny YM}}^2} \lambda_j ^2}}{2 \cosh \pi (\lambda_j + r_5 m)} ~,
\end{equation}
which equals the partition function of 3d $SU(N)$ Chern--Simons theory with one hypermultiplet, at imaginary Chern--Simons level. The matrix model is invariant under flipping $m \mapsto -m$, thus we restrict to $m \ge 0$ without loss of generality.\par
In the large $N$ limit at fixed gauge coupling, the defect free energy reads 
\begin{equation}
	F_{\cD} = \left. \sum_{j=1}^{N} \ln \left[2 \cosh \pi (\lambda_j +r_5 m) \right] \right\rvert_{\rm saddle} .
\end{equation}
The saddle points are \cite{Kallen:2012zn}
\begin{equation}
	\lambda_j \big\vert_{\rm saddle } = \frac{9 g_{\text{\tiny YM}}^2 }{64 \pi^2 r_5} \left( 2j - N\right) ~,
\end{equation}
and we get 
\begin{align}
	F_{\cD} = \frac{9 N^2}{64 \pi} \int_0 ^1 d \xi \left\lvert  \left( \frac{g_{\text{\tiny YM}}^2}{r_5} \right) (2 \xi -1) + r_5 \mu \right\rvert && m = \frac{9 N}{64 \pi^2} \mu ~,
\end{align}
which grows as $N^2$, subleading compared to the $N^3$ behaviour of the ambient free energy. Direct integration yields
\begin{equation}
	F_{\cD} =  \frac{9 N^2}{128 \pi} \times \begin{cases}  \frac{g_{\text{\tiny YM}}^2}{r_5} \left( 1 + \frac{\mu^2 r_5 ^3}{g_{\text{\tiny YM}}^2}\right) & \ 0 \le \mu r_5 < \frac{g_{\text{\tiny YM}}^2}{r_5} \\ 2 \mu r_5 & \ \mu r_5  > \frac{g_{\text{\tiny YM}}^2}{r_5} . \end{cases}
\end{equation}
This expression shows a second order phase transition as a function of the mass, with finite discontinuity at $ \mu r_5 = \frac{g_{\text{\tiny YM}}^2}{r_5}$. In the large mass phase, we simply have $F_{\cD} = N \pi m$, indicating that the defect contributes as a collection of $N$ free 3d massive hypermultiplets. This value is scheme-dependent and can be cancelled by a counterterm. In the small mass phase we have instead 
\begin{equation}
	F_{\cD}  = \frac{9 \pi r_6}{16 r_5 } N^2  + \frac{4 \pi r_5^3 m^2 }{9 r_6 } ~.
\end{equation}
The $\frac{r_6}{r_5} N^2$ behaviour of the first piece signals the origin of the defect as a 4d defect supported on $S^3 \times S^1$, and decreases monotonically as the ambient $\cN=2$ SYM theory flows to the IR. The polynomial dependence on $m^2$ is incompatible with a genuinely 3d system \cite{Closset:2012vg}, but is consistent with a 4d defect on $S^3 \times S^1$.

\bibliography{defect}
\end{document}